\documentclass[prodmode,acmcsur]{acmsmall} 

\usepackage[ruled]{algorithm2e}

\SetAlFnt{\small}
\SetAlCapFnt{\small}
\SetAlCapNameFnt{\small}
\SetAlCapHSkip{0pt}
\IncMargin{-\parindent}

\acmVolume{X}
\acmNumber{X}
\acmArticle{X}
\acmYear{2016}
\acmMonth{3}

\usepackage{lscape}
\usepackage{amsmath}
\usepackage{graphicx}
\usepackage{booktabs}
\usepackage{tabularx}
\usepackage{multirow}
\usepackage{stmaryrd}
\usepackage{subfigure}
\usepackage{url} 
\usepackage{color}
\usepackage{cite}
\usepackage{wasysym}
\usepackage{todonotes}
\usepackage{amssymb}
\usepackage{verbatim}

\newcommand{\rr}{\raggedright}

\newcommand{\hide}[1]{}

\newcommand{\etal}{\textit{et al.}~}

\newcommand{\specialcell}[2][l]{%
  \begin{tabular}[#1]{@{}l@{}}#2\end{tabular}}

\graphicspath{{./images/}}

\usepackage{framed}
\usepackage{color}

\newcommand{\red}[1]{\textcolor{red}{#1}}

\newcommand{\black}[1]{\textcolor{black}{#1}}
\newcommand{\tabincell}[2]{\begin{tabular}{@{}#1@{}}#2\end{tabular}}

\newenvironment{shadequote}%
{\begin{snugshade}\begin{quote}}
{\hfill\end{quote}\end{snugshade}}

\definecolor{shadecolor}{rgb}{0.9,0.9,0.9}

\makeatletter
\newcommand\footnoteref[1]{\protected@xdef\@thefnmark{\ref{#1}}\@footnotemark}
\makeatother

\begin{document}

\markboth{X. Li et al.}{Socializing the Semantic Gap}

\title{Socializing the Semantic Gap: A Comparative Survey on \\Image Tag Assignment, Refinement and Retrieval}
\author{XIRONG LI*
\affil{Renmin University of China} 
TIBERIO URICCHIO*
\affil{University of Florence}
LAMBERTO BALLAN
\affil{University of Florence, Stanford University}
MARCO BERTINI
\affil{University of Florence}
CEES G. M. SNOEK
\affil{University of Amsterdam, Qualcomm Research Netherlands}
ALBERTO DEL BIMBO
\affil{University of Florence}
}

\begin{abstract}
Where previous reviews on content-based image retrieval emphasize on what can be seen in an image to bridge the semantic gap, this survey considers what people tag about an image. A comprehensive treatise of three closely linked problems, i.e., image tag assignment, refinement, and tag-based image retrieval is presented. While existing works vary in terms of their targeted tasks and methodology, they rely on the key functionality of tag relevance, i.e. estimating the relevance of a specific tag with respect to the visual content of a given image and its social context. By analyzing what information a specific method exploits to construct its tag relevance function and how such information is exploited, this paper introduces a two-dimensional taxonomy to structure the growing literature, understand the ingredients of the main works, clarify their connections and difference, and recognize their merits and limitations. For a head-to-head comparison between the state-of-the-art, a new experimental protocol is presented, with training sets containing 10k, 100k and 1m images and an evaluation on three test sets, contributed by various research groups. Eleven representative works are implemented and evaluated. Putting all this together, the survey aims to provide an overview of the past and foster progress for the near future.
\end{abstract}




\acmformat{Xirong Li, Tiberio Uricchio, Lamberto Ballan, Marco Bertini, Cees G. M. Snoek, Alberto Del Bimbo, 2016. 
Socializing the Semantic Gap: A Comparative Survey on Image Tag Assignment, Refinement and Retrieval.}

\begin{bottomstuff}
* Equal contribution and corresponding authors. 
This research was supported by NSFC (No. 61303184), SRFDP (No. 20130004120006), the Fundamental Research Funds for the Central Universities and the Research Funds of Renmin University of China (No. 14XNLQ01, No. 16XNQ013), SRF for ROCS, SEM, the Dutch national program COMMIT, the STW STORY project, Telecom Italia PhD grant funds, and the AQUIS-CH project granted by the Tuscany Region (Italy). L. Ballan acknowledges also the support of the EC's FP7 under the grant agreement No. 623930 (Marie Curie IOF). \\
Author's addresses: 
X. Li, Key Lab of Data Engineering and Knowledge Engineering, School of Information, Renmin University of China;
C. Snoek, Intelligent Systems Lab Amsterdam, University of Amsterdam, Netherlands;
T. Uricchio, L. Ballan, M. Bertini, and A. Del Bimbo, Media Integration and Communication Center, University of Florence, Italy.
\end{bottomstuff}

\maketitle

\section{Introduction} \label{sec:intro}
%
Images want to be shared. Be it a drawing carved in rock, a painting exposed in a museum, or a photo capturing a special moment, it is the sharing that relives the experience stored in the image. Nowadays, several technological developments have spurred the sharing of images in unprecedented volumes. The first is the ease with which images can be captured in a digital format by cameras, cellphones and other wearable sensory devices. The second is the Internet that allows transfer of digital image content to anyone, anywhere in the world. Finally, and most recently, the sharing of digital imagery has reached new heights by the massive adoption of social network platforms. All of a sudden images come with tags. Tagging, commenting, and rating of any digital image has become a common habit. 
As a result, we observe a downpour of personally annotated user-generated visual content and associated metadata.
The problem of image retrieval has been dilated with the problem of searching images generated within social platforms and improving social media annotations in order to permit effective retrieval.

Excellent surveys on content-based image retrieval have been published in the past. In their seminal work, Smeulders \etal review the early years up to the year 2000 by focusing on what can be seen in an image and introducing the main scientific problem of the field: the semantic gap as ``the lack of coincidence between the information that one can extract from the visual data and the interpretation that the same data have for a user in a given situation" \cite{cbir-tpami00}. Datta \etal continue along this line and describe the coming-of-age of the field, highlighting the key theoretical and empirical contributions of recent years \cite{datta-cs2008}. These reviews completely ignore social platforms and socially generated images, which is not surprising as the phenomenon only became apparent after these reviews were published. 

In this paper, we survey the state-of-the-art of content-based image retrieval in the context of social image platforms and tagging, with a comprehensive treatise of the closely linked problems of image tag assignment, image tag refinement and tag-based image retrieval.
Similar to \cite{cbir-tpami00} and \cite{datta-cs2008}, the focus of our survey is on visual information, but we explicitly take into account \emph{and} quantify the value of social tagging.

\subsection{Problems and Tasks} \label{ssec:tasks}

Social image tags are provided by common users. They often cannot meet high quality standards related to content association, in particular for accurately describing objective aspects of the visual content according to some expert's opinion \cite{nnacl2012-dodge}. Social tags tend to follow context, trends and events in the real world. They are often used to describe both the situation and the entity represented in the visual content. 
In such a context there are distinct problems to solve. On the one hand, social tags tend to be imprecise, ambiguous and incomplete. On the other hand,  they are biased towards personal perspectives. So tagging deviations due to spatial and temporal correlation to external factors are common phenomena~\cite{Golder06,cscw2006-ssen,www2008-borkur,kennedy-2006}.
The focus of interests and motivations of an image retriever could be different from those of an image uploader.

Quite a few researchers have proposed solutions for image annotation and retrieval in social frameworks, 
although the peculiarities of this domain have been only partially addressed.
Concerning the role of visual content in social image tagging, 
several studies have shown that people are willing to tag objects and scenes presented in the visual content to favor image retrieval for general audience \cite{Ames-2007,www2008-borkur,Nov-2010}.
It would be relevant to survey why people search images on social media platforms and what query terms they actually use. 
Although some query log data of generic web image search have been made publicly accessible \cite{mm2013-hua}, its social-media counterpart remains to be established. 
Most of the existing works have rather investigated the technological possibilities to automatically assign, refine, and enrich image tags. 
They mainly concentrated on how to expand the set of tags provided by the uploader, by looking at tags that others have associated to similar content, so expecting to include tags suited to the retriever's motivations. 
Consequently, images will become findable and potentially appreciated by a wider range of audiences beyond the relatively small social circle of the image uploader. 
We categorize these existing works into three different main tasks and structure our survey along these tasks:



%
\begin{itemize} 
  \item \textbf{Tag Assignment}. 
  Given an unlabeled image, tag assignment strives to assign a (fixed) number of tags related to the image content \cite{ijcv10-makadia,iccv09-tagprop,mir2010-verbeek,tist11-tang}. 
  \medskip
  \item \textbf{Tag Refinement}. 
 Given an image associated with some initial tags, tag refinement aims to remove irrelevant tags from the initial tag list and enrich it with novel, yet relevant, tags \cite{mm2010-liuretagging,pami2013-wutag,icmr2013-aznaidia,cvpr2013-zlin,eccv2014-zfeng}.
  \medskip
  \item \textbf{Tag Retrieval}. Given a tag and a collection of images labeled with the tag (and possibly other tags), the goal of tag retrieval is to retrieve images relevant with respect to the tag of interest~\cite{tmm09-xirong,tip11-duan,jasist11-sun,tip13-gao,pami2013-wutag}. 
  \medskip
\end{itemize}
%
 
Other related tasks such as tag filtering \cite{mm10-zhu,tmm11-liu,tmm12-zhu} and tag suggestion \cite{www2008-borkur,tmm09-xirong,www09-lwu} have also been studied. 
We view them as variants of tag refinement.

As a common factor in all the works for tag assignment, refinement and retrieval, we reckon that the way in which the tag set expansion is performed relies on the key functionality of \emph{tag relevance}, i.e., estimating the relevance of a tag with respect to the visual content of a given image and its social context.

\subsection{Scope, Aims, and Organization} \label{ssec:scope}


We survey papers that learn tag relevance from images tagged in social contexts.
While it would have been important to consider the complementarity of tags, only a few methods have considered multi-tag retrieval \cite{tmm12-li,mm2012-nie,mm2013-borth}.
Hence, we focus on methods that implement the unique-tag relevance model. 
We do not cover traditional image classification that is grounded on carefully labeled data. 
For a state-of-the-art overview in that direction, we refer the interested reader to \cite{everingham-ijcv-2014,ILSVRCarxiv14}. Nonetheless, one may question the necessity of using socially tagged examples as training data, given that a number of labeled resources are already publicly accessible. 
An exemplar of such resources is ImageNet \cite{cvpr2009-imagenet}, providing crowd-sourced positive examples for over 20k classes. 
Since ImageNet employs several web image search engines to obtain candidate images,
its positive examples tend to be biased by the search results.
As observed by \cite{icmr2012-vreeswijk}, the positive set of vehicles mainly consists of car and buses, although vehicles can be tracks, watercraft and aircraft.
Moreover, controversial images are discarded upon vote disagreement during the crowd sourcing.
All this reduces diversity in visual appearance.
We empirically show in Section \ref{ssec:imagenet} the advantage of socially tagged examples against ImageNet for tag relevance learning.

Reviews on social tagging exist. 
The work by Gupta \etal discusses papers on why people tag, what influences the choice of tags, and how to model the tagging process, but its discussion on content-based image tagging is limited \cite{kdd2010-gupta}.
The focus of \cite{mtap2015-jabeen} is on papers about adding semantics to tags by exploiting varied knowledge sources such as Wikipedia, DBpedia, and WordNet. Again, it leaves the visual information untouched.


Several reviews that consider socially tagged images have appeared recently.
In \cite{mtap11-liu}, technical achievements in content-based tag processing for social images are briefly surveyed. 
Sawant \etal \cite{mtap11-sawant}, Wang \etal \cite{csur2012-wang} and Mei \etal \cite{csur2014-mei} present extended reviews of particular aspects, i.e., collaborative media annotation, assistive tagging, and visual search re-ranking, respectively. 
In \cite{mtap11-sawant}, papers that propose collaborative image labeling games and tagging in social media networks are reviewed. 
In \cite{csur2012-wang} the authors survey papers where computers assist humans in tagging either by organizing data for manual labelling, improving quality of human-provided tags or recommending tags for manual selection, instead of applying purely automatic tagging. 
In \cite{csur2014-mei} the authors review techniques that aim for improving initial search results, typically returned by a text based visual search engine, by visual search re-ranking.
These reviews offer resumes of the methods and interesting insights on particular aspects of the domain,
without giving an experimental comparison between the varied methods.

We notice efforts in empirical evaluations of social media annotation and retrieval \cite{jasist11-sun,icme13-uricchio,ballan2014data}. 
In \cite{jasist11-sun}, the authors analyze different dimensions to compute the relevance score between a tagged image and a tag. They evaluate varied combinations of these dimensions for tag-based image retrieval on NUS-WIDE,
a leading benchmark set for social image retrieval \cite{civr09-chua}.
However, their evaluation focuses only on tag-based image ranking features,
without comparing content-based methods.
Moreover, tag assignment and refinement are not covered.
In \cite{icme13-uricchio,ballan2014data}, the authors compared three algorithms for tag refinement on the NUS-WIDE and 
MIRFlickr, a popular benchmark set for tag assignment and refinement \cite{mir10-huiskes}.
However, the two reviews lack a thorough comparison between different methods under the umbrella of a common experimental protocol. 
Moreover, they fail to assess the high-level connection between image tag assignment, refinement, and retrieval.

The aims of this survey are twofold. First, we organize the rich literature in a taxonomy to highlight the ingredients of the main works in the literature and recognize their advantages and limitations. 
In particular, we structure our survey along the line of understanding how a specific method constructs the underlying tag relevance function.
Witnessing the absence of a thorough empirical comparison in the literature, our second goal is to establish a common experimental protocol and successively exert it in the evaluation of key methods. Our proposed protocol contains training data of varied scales extracted from social frameworks. This permits to evaluate the methods under analysis with data that reflect the specificity of the social domain. 
We have made the data and source code public\footnote{\label{codeurl}\url{https://github.com/li-xirong/jingwei}} so that new proposals for tag assignment, tag refinement, and tag retrieval can be evaluated rigorously and easily. 
Taken together, these efforts should provide an overview of the field's past and foster progress for the near future.

The rest of the survey is organized as follows. Section \ref{sec:main} introduces a taxonomy to structure the literature on tag relevance learning. 
Section \ref{sec:protocol} proposes a new experimental protocol for evaluating the three tasks. A selected set of eleven representative works, described in Section \ref{sec:selected}, is compared extensively using this protocol, with results and analysis provided in Section \ref{sec:eval}. We provide concluding remarks and our vision about future directions in Section \ref{sec:conclusions}.




\section{Taxonomy and Review} \label{sec:main}
\subsection{Foundations} \label{ssec:founda}

Our key observation is that the essential component, which measures the relevance between a given image and a specific tag, stands at the heart of the three tasks. In order to describe this component in a more formal way, we first introduce some notation.

We use $x$, $t$, and $u$ to represent three basic elements in social images, namely image, tag, and user.
An image $x$ is shared on social media by its user $u$. 
A user $u$ can choose a specific tag $t$ to label $x$.
By sharing and tagging images, a set of users $\mathcal{U}$ contribute a set of $n$ socially tagged images $\mathcal{X}$,
wherein $\mathcal{X}_t$ denotes the set of images tagged with $t$.
Tags used to describe the image set form a vocabulary of $m$ tags $\mathcal{V}$.
The relationship between images and tags can be represented by an image-tag association matrix 
$D\in \{0, 1\}^{n \times m}$, where $D_{ij}=1$ means the $i$-th image is labeled with the $j$-th tag, and 0 otherwise.

Given an image and a tag, we introduce a real-valued function 
that computes the relevance between $x$ and $t$ based on the visual content and an optional set of user information $\Theta$ associated with the image:
\begin{shadequote}
$$
f_\Phi(x,t ; \Theta)
$$
\end{shadequote}
We use $\Theta$ in a broad sense, making it refer to any type of social context provided by or referring to the user like 
associated tags, where and when the image was taken, personal profile, and contacts.
The subscript $\Phi$ specifies how the tag relevance function is constructed. 

Having $f_\Phi(x,t ; \Theta)$ defined, 
we can easily interpret each of the three tasks.
Assignment and refinement can be done by sorting $\mathcal{V}$ in descending order by $f_\Phi(x,t ; \Theta)$, while retrieval can be achieved by sorting the labeled image set $\mathcal{X}_t$ in descending order in terms of $f_\Phi(x,t ; \Theta)$.
Note that this formalization does not necessarily imply that the same implementation of tag relevance is applied for all the three tasks. For example, for retrieval relevance is intended to obtain image ranking \cite{msys2014-li} while tag ranking for each single image is the goal of assignment \cite{www09-lwu} and refinement \cite{tcyb2014-qian}.

\begin{figure*}[!tb]
\centering
\noindent\includegraphics[width=\columnwidth]{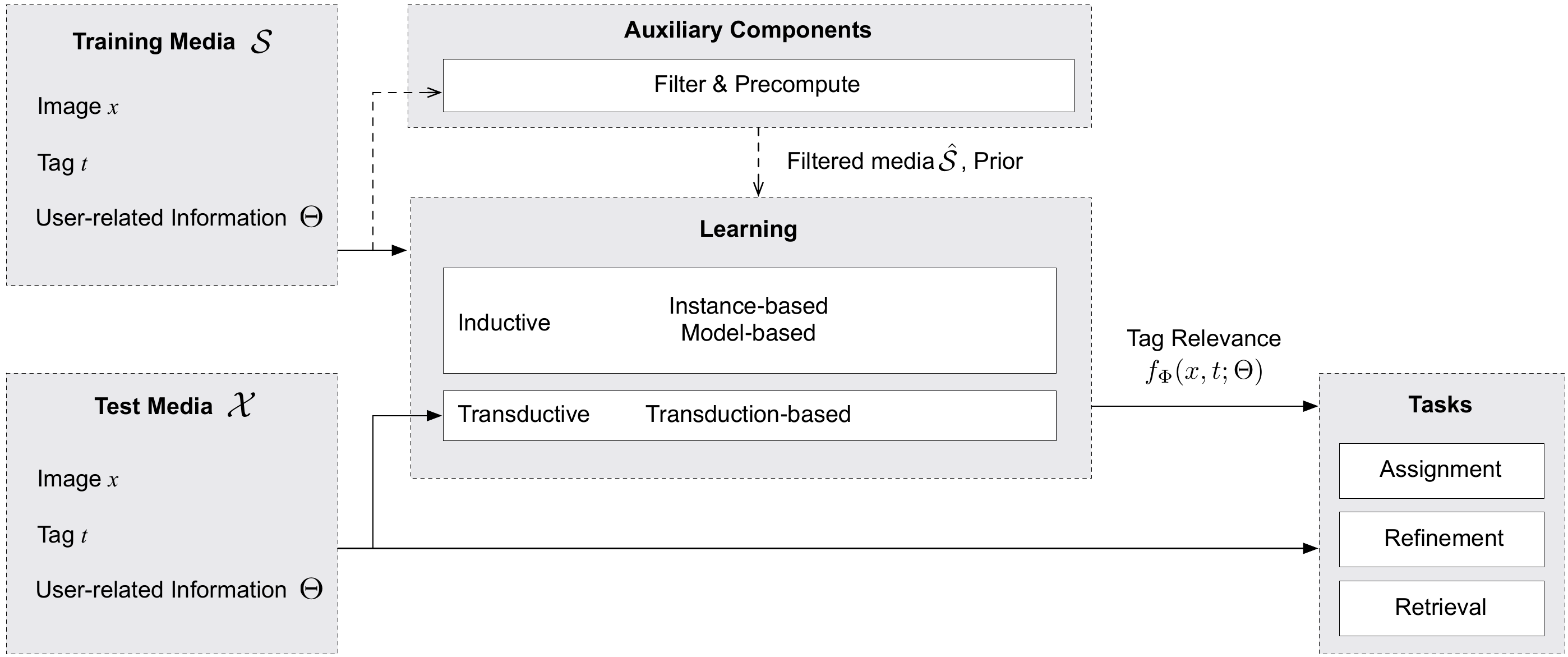}
\caption{\textbf{Unified framework of tag relevance learning for image tag assignment, refinement and retrieval.} 
We follow the input data as it flows through the process of learning the tag relevance funciton $f_\Phi(x,t;\Theta)$ to higher-level tasks.
Dashed lines indicate optional data flow.
The framework jointly classifies existing works on Assignment, Refinement and Retrieval while at the same determining their main components.}\label{fig:pipeline}
\end{figure*}

Fig. \ref{fig:pipeline} presents a unified framework, illustrating the main data flow of varied approaches to tag relevance learning.
Compared to traditional methods that rely on expert-labeled examples,
a novel characteristic of a social media based method is its capability to learn from socially tagged examples with unreliable and personalized annotations. Such a training media is marked as $\mathcal{S}$ in the framework and includes tags, images or user-related information.
Optionally, in order to obtain a refined training media $\mathcal{\hat{S}}$, 
one might consider designing a filter to remove unwanted data.
In addition, prior information such as tag statistics, tag correlations, and image affinities in the training media are independent of a specific image-tag pair.
They can be precomputed for the sake of efficiency.
As the filter and the precomputation appear to be a choice of implementation, they are positioned as auxiliary components in Fig. \ref{fig:pipeline}.

A number of implementations of the relevance function have been proposed that utilizes different modes to expand the tag set by learning within the social context. 
They may exploit different media, such as tags only, tags and related image content, or tags, image content and user-related information.
Depending on how $f_\Phi(x,t;\Theta)$ is composed internally, we propose a taxonomy which organizes existing works along two dimensions,
namely \emph{media} and \emph{learning}.
The media dimension characterizes \emph{what} essential information $f_\Phi(x,t;\Theta)$  exploits,
while the learning dimension depicts \emph{how} such information is exploited. 
Table \ref{tab:taxonomy} presents a list of the most significant contributions organized along these two dimensions.
For a specific work, while Fig. \ref{fig:pipeline} helps illustrate the main data flow of the method,
its position in the two-dimensional taxonomy is pinpointed via Table \ref{tab:taxonomy}.
We believe such a context provides a good starting point for an in-depth understanding of the work.
We explore the taxonomy along the media dimension in Section \ref{ssec:media} 
and the learning dimension in Section  \ref{ssec:learning}.
Auxiliary components are addressed in Section \ref{ssec:auxiliary}. 
A comparative evaluation of a few representative methods is presented in Section \ref{sec:selected}.

\subsection{Media for tag relevance} \label{ssec:media}

Different sources of information may play a role in determining the relevance between an image and a social tag.
For instance, the position of a tag appearing in the tag list might reflect a user's tagging priority to some extent \cite{jasist11-sun}.
Knowing what other tags are assigned to the image \cite{tmm12-zhu} or what other users label about similar images \cite{tmm09-xirong,kennedy-2009} can also be helpful for judging whether the tag under examination is appropriate or not. 
Depending on what modalities in $\mathcal{S}$ are utilized, 
we divide existing works into the following three groups:
\textit{1)} tag based, \textit{2)} tag $+$ image based and \textit{3)} tag $+$ image  $+$ user information based,
ordered in light of the amount of information they utilize. Table \ref{tab:taxonomy} shows this classification for several papers that appeared in the literature on the subject.

\subsubsection{Tag based}
These methods build $f_\Phi(x,t ; \Theta)$ purely based on tag information. 
The basic idea is to assign higher relevance scores to tags that are semantically close to the majority of the tags associated with the test image.
To that end, 
in \cite{www2008-borkur,tmm12-zhu} relevant tags are suggested based on tag co-occurrence statistics mined from large-scale collections, 
while topic modeling is employed in \cite{mm09-hxu}.
As the tag based methods presume that the test image has been labeled with some initial tags, 
i.e.~the initial tags are taken as the user information $\Theta$,
they are inapplicable for tag assignment.

\subsubsection{Tag $+$ Image based}
Works in this group develop $f_\Phi(x,t ; \Theta)$ on the base of visual information and associated tags. 
The main rationale behind them is visual consistency, i.e.~visually similar images shall be labeled with similar tags.
Implementations of this intuition can be grouped in three conducts.
One, leverage images visually close to the test image \cite{tmm09-xirong,civr10-xirong,mir2010-verbeek,tmm2010-ma,wsdm2011-pwu,mm12-feng}. 
Two, exploit relationships between images labeled with the same tag \cite{www09-liu,mtap2010-richter,tmm11-liu,tmm12-kuo,tip13-gao}. 
Three, learn visual classifiers from socially tagged examples \cite{cvpr2009-wang,tmm12-chen,mm13-xirong,icmr2014-yyang}. 
By propagating tags based on the visual evidence, the above works exploit the image modality and the tag modality in a sequential way.
By contrast, there are works that concurrently exploit the two modalities. 
This can be approached by generating a common latent space upon the image-tag association \cite{nips2012-srivastava,cvpr2014-niu,cvpr2014-duan}, so that a cross media similarity can be computed between images and tags \cite{wsdm2011-zhuang,tpami2012-gqi,neurocom2013-jliu}.
In \cite{pami2014-pereira}, the latent space is constructed by Canonical Correlation Analysis,
finding two matrices which separately project feature vectors of image and tag into the same subspace.
In \cite{tmm2010-ma}, a random walk model is used on a unified graph composed from the fusion of an image similarity graph with an image-tag connection graph.
In \cite{pami2013-wutag,icme2014-xxu,mm10-zhu}, predefined image similarity and tag similarity are used as two constraint terms to enforce that similarities induced from the recovered image-tag association matrix will be consistent with the two predefined similarities.

Although late fusion has been actively studied for multimedia data analysis \cite{multimodal-fusion-survey10},
improving tag relevance estimation by late fusion is not much explored. 
There are some efforts in that direction,
among which interesting performance has been reported in \cite{tcyb2014-qian} 
and more recently in \cite{msys2014-li}. 


\subsubsection{Tag $+$ Image $+$ User-related information based}
In addition to tags and images, this group of works exploit user information, motivated from varied perspectives. 
User information ranges from the simplest user identities \cite{tmm09-xirong}, tagging preferences \cite{icmr2010-sawant} to user reliability \cite{mm2014-ginsca} and to image group memberships \cite{iccv2015-ballan}.
With the hypothesis that a specific tag chosen by many users to label visually similar images is more likely to be relevant with respect to the visual content,
\cite{tmm09-xirong} utilizes user identity to ensure that learning examples come from distinct users.
A similar idea is reported in \cite{kennedy-2009}, finding visually similar image pairs with matching tags from different users.
\cite{mm2014-ginsca} improves image retrieval by favoring images uploaded by users with good credibility estimates.
The reliability of an image uploader is inferred by counting matches between the user-provided tags and machine tags predicted by visual concept detectors.
In \cite{icmr2010-sawant,mm2011-pia}, personal tagging preference is considered in the form of tag statistics computed from images a user has uploaded in the past.
These past images are used in \cite{tmm2014-jliu} to learn a user-specific embedding space.
In \cite{tmm12-sang}, user affinities, measured in terms of the number of common groups users are sharing, is considered in a tensor analysis framework.
Similarly, tensor based low-rank data reconstruction is employed in \cite{spl2015-qian} to discover latent associations between users, images, and tags.
Photo timestamps are exploited for time-sensitive image retrieval \cite{kim2013-time},
where the connection between image occurrence and various temporal factors is modeled.
In \cite{mmm2013-mcparlane}, time-constrained tag co-occurrence statistics are considered to refine the output of visual classifiers for tag assignment.
In their follow-up work \cite{sigir2013-mcparlane}, 
location-constrained tag co-occurrence computed from images taken in a specific continent is further included.
User interactions in social networks are exploited in \cite{icmr2010-sawant}, computing local interaction networks from the comments left by other users.
In \cite{eccv2012-mcauley,iccv2015-ballan},
social-network metadata such as image groups membership or contacts of users is employed to resolve ambiguity in visual appearance.

\medskip
Comparing the three groups, tag + image appears to be the mainstream, as evidenced by the imbalanced distribution in Table \ref{tab:taxonomy}.
Intuitively, using more media from $\mathcal{S}$ would typically improve tag relevance estimation.
We attribute the imbalance among the groups, in particular the relatively few works in the third group, to the following two reasons.
First, no publicly available dataset with expert annotations was built to gather representative and adequate user information, e.g.~MIRFlickr has nearly 10k users for 25k images, while in NUS-WIDE only 6\% of the users have at least 15 images. 
As a consequence, current works that leverage user information are forced to use a minimal subset to alleviate sample insufficiency \cite{tmm12-sang,tmm12-dlu} or homemade collections with social tags as ground truth instead of benchmark sets \cite{icmr2010-sawant,mm2011-pia}.
Second, adding more media often results in a substantial increase in terms of both computation and memory,
e.g.~the cubic complexity for tensor factorization in \cite{tmm12-sang}.
As a trade-off, one has to use $\mathcal{S}$ of a much smaller scale.
The dilemma is whether one should use large data with less media or more media but less data.

It is worth noting that  the above groups are not exclusive. The output of some methods can be used as a refined input of some other methods. In particular, we observe a frequent usage of tag-based methods by others for their computational efficiency.
For instance, tag relevance measured in terms of tag similarity is used in \cite{wsdm2011-zhuang,tip13-gao,mm13-xirong} before applying more advanced analysis, while nearest neighbor tag propagation is a pre-process used in \cite{mm10-zhu}.
The number of tags per image is embedded into image retrieval functions in \cite{www09-liu,mm09-hxu,wsdm2011-zhuang,tmm12-chen}.

Given the varied sources of information one could leverage, the subsequent question is how the information is exactly utilized, which will be made clear next.

\begin{table} [tb!]
\renewcommand{\arraystretch}{1.3}
\tbl{The taxonomy of methods for tag relevance learning, organized along the \emph{Media} and \emph{Learning} dimensions of Fig. \ref{fig:pipeline}. Methods for which this survey provides an experimental evaluation are indicated in \textbf{bold font}.\label{tab:taxonomy}}
{\centering
\scalebox{0.9}{
\begin{tabular}{@{}l l l l @{}}
\toprule

   & \multicolumn{3}{c}{\textbf{Learning}} \\
\cmidrule{2-4}         
\textbf{Media}      &  \textit{Instance-based} & \textit{Model-based} & \textit{Transduction-based} \\
\cmidrule{1-4}  

\textit{tag}  & \specialcell{\textbf{\cite{www2008-borkur}} \\  \textbf{\cite{tmm12-zhu}} }  & \cite{mm09-hxu} & --\\

\cmidrule{2-4} 

\textit{tag + image} & \specialcell{\textbf{\cite{www09-liu}}  \\ \textbf{\cite{ijcv10-makadia}} \\ \cite{tist11-tang} \\ \cite{wsdm2011-pwu} \\ \cite{tmm11-yang} \\ \cite{icmr2012-truong} \\ \cite{tpami2012-gqi}  \\ \cite{cvpr2013-zlin} \\ \cite{mtap2013-lee} \\ \cite{icme13-uricchio}  \\ \cite{sigir2014-xzhu} \\ \cite{ballan2014-icmr} \\ \cite{pami2014-pereira} } & \specialcell{\cite{www09-lwu} \\ \textbf{\cite{iccv09-tagprop}} \\ \cite{mir2010-verbeek} \\ \cite{mm2010-liuretagging} \\ \cite{tmm2010-ma} \\ \cite{tmm11-liu} \\ \cite{tip11-duan} \\  \cite{mm12-feng} \\ \cite{nips2012-srivastava} \\ \textbf{\cite{tmm12-chen}} \\ \cite{cvpr2013-tlan}    \\ \textbf{\cite{mm13-xirong}} \\ \cite{neurcomp2013-li} \\ \cite{cviu2014-ldatag} \\ \cite{cvpr2014-niu}} & \specialcell{ \textbf{\cite{mm10-zhu}}  \\ \cite{tmm10-wang}  \\ \cite{mm2010-zli}  \\ \cite{wsdm2011-zhuang} \\   \cite{mtap2010-richter} \\ \cite{tmm12-kuo} \\ \cite{neurocom2013-jliu}  \\ \cite{tip13-gao} \\ \cite{pami2013-wutag}  \\ \cite{icmr2014-yyang} \\ \cite{eccv2014-zfeng} \\\cite{icme2014-xxu} } \\

\cmidrule{2-4} 

\textit{tag + image + user} & \specialcell{\textbf{\cite{tmm09-xirong}} \\ \cite{kennedy-2009} \\ \cite{civr10-xirong} \\ \cite{icmr2013-aznaidia} \\ \cite{tmm2014-jliu} }  & \specialcell{ \cite{icmr2010-sawant} \\  \cite{mm2011-pia}\\ \cite{eccv2012-mcauley} \\  \cite{kim2013-time} \\ \cite{sigir2013-mcparlane} \\ \cite{mm2014-ginsca} \\ \cite{iccv2015-ballan} } & \specialcell{ \textbf{\cite{tmm12-sang}} \\ \cite{tmm12-dlu} \\ \cite{spl2015-qian}} \\

\bottomrule

\end{tabular}
}}
\end{table}

\subsection{Learning for tag relevance} \label{ssec:learning}

This section presents the second dimension of the taxonomy, elaborating on various algorithms that implements the computation of tag relevance. Ideally, given the large-scale nature of social images, a desirable algorithm shall maintain a good scalability as the data grows. The algorithm shall also provide a flexible mechanism to effectively integrate various types of information including tags, images, social metadata, etc, while at the same time, being robust when not all the information is available. In what follows we review existing algorithms on their efforts to meet the requirements.

Depending on whether the tag relevance learning process is transductive, i.e., producing tag relevance scores without distinction as training and testing, we divide existing works into transduction-based and induction-based.
Since the latter produces rules or models that are directly applicable to a novel instance \cite{michalski1983theory},
it has a better scalability for large-scale data compared to its transductive counterpart.
Depending on whether an explicit model, let it be discriminative or generative, is built,
a further division for the induction-based methods can be made: instance-based algorithms and model-based algorithms.
Consequently, we divide existing works into the following three exclusive groups: 
\textit{1)} instance-based, \textit{2)} model-based, and \textit{3)} transduction-based.

\subsubsection{Instance-based}
This class of methods does not perform explicit generalization but, instead, compares new test images with training instances. 
It is called instance-based because it constructs hypotheses directly from the training instances themselves. 
These methods are nonparametric and the complexity of the learned hypotheses grows as the amount of training data increases. 
The neighbor voting algorithm \cite{tmm09-xirong} and its variants \cite{kennedy-2009,civr10-xirong,icmr2012-truong,mtap2013-lee,sigir2014-xzhu} estimate the relevance of a tag $t$ with respect to an image $x$ by counting the occurrence of $t$ in annotations of the visual neighbors of $x$.
The visual neighborhood is created using features obtained from early-fusion of global features \cite{tmm09-xirong}, 
distance metric learning to combine local and global features \cite{mir2010-verbeek,wsdm2011-pwu}, 
cross modal learning of tags and image features \cite{tpami2012-gqi,ballan2014-icmr,pami2014-pereira}, 
and fusion of multiple single-feature learners \cite{civr10-xirong,msys2014-li}.
While the standard neighbor voting algorithm \cite{tmm09-xirong} simply let the neighbors vote equally, 
efforts have been made to (heuristically) weight neighbors in terms of their importance.
For instance, in \cite{icmr2012-truong,mtap2013-lee} the visual similarity is used as the weights.
As an alternative to such a heuristic strategy,
\cite{sigir2014-xzhu} models the relationships among the neighbors by constructing a directed voting graph,
wherein there is a directed edge from image $x_i$ to image $x_j$ if $x_i$ is in the $k$ nearest neighbors of $x_j$.
Subsequently an adaptive random walk is conducted over the voting graph to estimate the tag relevance.
However, the performance gain obtained by these weighting strategies appears to be limited \cite{sigir2014-xzhu}.
The kernel density estimation technique used in \cite{www09-liu} can be viewed as another form of weighted voting, but the votes come from images labeled with $t$ instead of the visual neighbors. 
\cite{tmm11-yang} further considers the distance of the test image to images not labeled with $t$.
In order to eliminate semantically unrelated samples in the neighborhood,
sparse reconstruction from a $k$-nearest neighborhood is used in \cite{mm2009-tang,tist11-tang}.
In \cite{cvpr2013-zlin}, 
with intention of recovering missing tags by matrix reconstruction, the image and tag modalities are separately exploited in parallel to produce a new candidate image-tag association matrix each. 
Then, the two resultant tag relevance scores are linearly combined to produce the final tag relevance scores. 
To address the incompleteness of tags associated with the visual neighbors,
\cite{icmr2013-aznaidia} proposes to enrich these tags by exploiting tag co-occurrence in advance to neighbor voting.

\subsubsection{Model-based}
This class of tag relevance learning algorithms puts their foundations on parameterized models learned from the training media.
Notice that the models can be tag-specific or holistic for all tags.
As an example of holistic modeling,
a topic model approach is presented in \cite{cviu2014-ldatag} for tag refinement,
where a hidden topic layer is introduced between images and tags.
Consequently, the tag relevance function is implemented as the dot product between the topic vector of the image and the topic vector of the tag.
In particular, the authors extend the Latent Dirichlet Allocation model \cite{jmlr2003-lda} to force images with similar visual content to have similar topic distribution. According to their experiments \cite{cviu2014-ldatag}, however, the gain of such a regularization appears to be marginal compared to the standard Latent Dirichlet Allocation model.
\cite{neurcomp2013-li} first finds embedding vectors of training images and tags using the image-tag association matrix of $\mathcal{S}$.
The embedding vector of a test image is obtained by a convex combination of the embedding vectors of its neighbors retrieved in the original visual feature space.
Consequently, the relevance score is computed in terms of the Euclidean distance between the embedding vectors of the test image and the tag.

For tag-specific modeling,
linear SVM classifiers trained on features augmented by pre-trained classifiers of popular tags are used in \cite{tmm12-chen} for tag retrieval. 
Fast intersection kernel SVMs trained on selected relevant positive and negative examples are used in \cite{mm13-xirong}. 
A bag-based image reranking framework is introduced in \cite{tip11-duan},
where pseudo relevant images retrieved by tag matching are partitioned into clusters using visual and textual features.
Then, by treating each cluster as a bag and images within the cluster as its instances,
multiple instance learning \cite{nips03-mil} is employed to learn multiple-instance SVMs per tag.
Viewing the social tags of a test image as ground truth, 
a multi-modal tag suggestion method based on both tags and visual correlation is introduced in \cite{www09-lwu}.
Each modality is used to generate a ranking feature, and the tag relevance function is a combination of these ranking features, 
with the combination weights learned online by the RankBoost algorithm \cite{rankboost-jmlr03}.
In \cite{iccv09-tagprop,mir2010-verbeek}, logistic regression models are built per tag to promote rare tags.
In a similar spirit to \cite{mm13-xirong},
\cite{cvpr2015-zhou} learns an ensemble of SVMs by treating tagged images as positive training examples
and untagged images as candidate negative training examples.
Using the ensemble to classify image regions generated by automated image segmentation,
the authors assign tags at the image level and the region level simultaneously.



\subsubsection{Transduction-based}
This class of methods consists in procedures that evaluate tag relevance for all image-tag pairs by minimizing a specific cost function. 
Given the initial image-tag association matrix $D$, the output of the procedures is a new matrix $\hat{D}$ the elements of which are taken as tag relevance scores. Due to this formulation, no explicit form of the tag relevance function exists nor any distinction between training and test sets \cite{icml1999-joachims}. If novel images are added to the initial set,  minimization of the cost function needs to be re-computed.

The majority of transduction-based approaches are founded on matrix factorization \cite{mm10-zhu,tmm12-sang,neurocom2013-jliu,pami2013-wutag,cvpr2014-shah,eccv2014-zfeng,icme2014-xxu}.
In \cite{wsdm2011-zhuang} the objective function is a linear combination of the difference between $\hat{D}$ and the matrix of image similarity,
the distortion between $\hat{D}$ and the matrix of tag similarity, and the difference between $\hat{D}$ and $D$.
A stochastic coordinate descent optimization is applied to a randomly chosen row of $\hat{D}$  per iteration.
In \cite{mm10-zhu},
considering the fact that $D$ is corrupted with noise derived by missing or over-personalized tags, 
robust principal component analysis with laplacian regularization is applied to recover $\hat{D}$ as a low-rank matrix.
%
In \cite{pami2013-wutag}, $\hat{D}$ is regularized such that the image similarity induced from $\hat{D}$ is consistent with the image similarity computed in terms of low-level visual features, and the tag similarity induced from $\hat{D}$ is consistent with the tag correlation score computed in terms of tag co-occurrence. 
\cite{icme2014-xxu} proposes to re-weight the penalty term of each image-tag pair by their relevance score, 
which is estimated by a linear fusion of tag-based and content-based relevance scores.
To incorporate the user element, \cite{tmm12-sang} extends $D$ to a three-way tensor with tag, image, and user as each of the ways.
A core tensor and three matrices representing the three media, obtained by Tucker decomposition \cite{psych1966-tucker}, are multiplied to construct $\hat{D}$. 

As an alternative approach, in \cite{eccv2014-zfeng} it is assumed that the tags of an image are drawn independently from a fixed but unknown multinomial distribution.
Estimation of this distribution is implemented by maximum likelihood with low-rank matrix recovery and laplacian regularization like \cite{mm10-zhu}.

Graph-based label propagation is another type of transduction-based methods. 
In \cite{mtap2010-richter,tmm10-wang,tmm12-kuo}, the image-tag pairs are represented as a graph in which each node corresponds to a specific image and the edges are weighted according to a multi-modal similarity measure. 
Viewing the top ranked examples in the initial search results as positive instances,
tag refinement is implemented as a semi-supervised labeling process by propagating labels from the positive instances to the remaining examples using random walk.
While the edge weights are fixed in the above works,
\cite{tip13-gao} argues that fixing the weights could be problematic,
because tags found to be discriminative in the learning process should adaptively contribute more to the edge weights.
In that regard, the hypergraph learning algorithm \cite{nips2007-dzhou} is exploited and weights are optimized by minimizing a joint loss function which considers both the graph structure and the divergence between the initial labels and the learned labels.
In \cite{icdm2011-liu}, the hypergraph is embedded into a lower-dimension space by hypergraph Laplacian.



Comparing the three groups of methods for learning tag relevance,
an advantage of instance-based methods against the other two groups is their flexibility to adapt to previously unseen images and tags. 
They may simply add new training images into $\mathcal{S}$ or remove outdated ones.
The advantage however comes with a price that $\mathcal{S}$ has to be maintained, a non-trivial task given the increasing amount of training data available. Also, the computational complexity and memory footprint grow linearly with respect to the size of $\mathcal{S}$.
In contrast, model-based methods could be more swift, especially when linear classifiers are used, 
as the training data is compactly represented by a fixed number of models.
As the imagery of a given tag may evolve, re-training is required to keep the models up-to-date.

Different from instance-based and model-based learning where individual tags are considered independently, 
transduction-based learning methods via matrix factorization can favorably exploit inter-tag and inter-image relationships.
However, their ability to deal with the extremely large number of social images is a concern. 
For instance, the use of Laplacian graphs results in a memory complexity of $O(|\mathcal{S}|^2)$.  
The accelerated proximal gradient algorithm used in \cite{mm10-zhu} requires Singular Value Decomposition,
which is known to be an expensive operation.
The Tucker decomposition used in  \cite{tmm12-sang}  has a cubic computational complexity with respect to the number of training samples.
We notice that some engineering tricks have been considered in these works, which alleviate the scalability issue to some extent.
In \cite{wsdm2011-zhuang}, for instance, clustering is conducted in advance to divide $\mathcal{S}$ into much smaller subsets, 
and the algorithm is applied to these subsets, separately. By making the Laplacian more sparse by retaining only the $k$ nearest neighbors \cite{mm10-zhu,tmm12-sang}, the memory footprint can be reduced to $O(k\cdot |\mathcal{S}|)$, with the cost of performance degeneration. Perhaps due to the scalability concern, works resorting to matrix factorization tend to experiment with a dataset of relatively small scale.


In summary, instance-based learning, in particular neighbor voting, is the first choice to try for its simplicity and decent performance. When the test tags are well defined (in the sense of relevant learning examples that can be collected automatically), model-based learning is more attractive. When the test images share similar social context, e.g., images shared by a group of specific interest, they tend to be on similar topics. In such a scenario, transduction-based learning that exploits the inter-image relationship is more suited.

\subsection{Auxiliary components} \label{ssec:auxiliary}

The \textit{Filter} and the \textit{Precompute} component are auxiliary components that may sustain and improve tag relevance learning.

\textit{Filter.} As social tags are known to be subjective and overly personalized, 
removing personalized tags appears to be a natural and simple way to improve the tagging quality. 
This is usually the first step performed in the framework for tag relevance learning.
Although there is a lack of golden criteria to determine which tags are personalized,
a popular strategy is to exclude tags which cannot be found in the WordNet ontology \cite{mm10-zhu,mm2011-pia,tmm12-chen,tmm12-zhu} or a Wikipedia thesaurus \cite{www09-liu}. Tags with rare occurrence, say appearing less than 50 times, are discarded in \cite{mir2010-verbeek,mm10-zhu}. For methods that directly work on the image-tag association matrix \cite{mm10-zhu,tmm12-sang,pami2013-wutag,cvpr2013-zlin}, reducing the size of the vocabulary in terms of tag occurrence is an important prerequisite to keep the matrix in a manageable scale. Observing that images tagged in a batch manner are often nearly duplicate and of low tagging quality, batch-tagged images are excluded in \cite{tmm12-li}. Since relevant tags may be missing from user annotations, the negative tags that are semantically similar or co-occurring with positive ones are discarded in \cite{tmm12-sang}. As the above strategies do not take the visual content into account, they cannot handle situations where an image is incorrectly labeled with a valid and frequently used tag, say `dog'. In \cite{icassp2009-xirong}, tag relevance scores are assigned to each image in $\mathcal{S}$ by running the neighbor voting algorithm \cite{tmm09-xirong}, while in \cite{mm13-xirong}, the semantic field algorithm \cite{tmm12-zhu} is further added to select relevant training examples.
In \cite{spl2015-qian}, the annotation of the training media is enriched by a random walk.

\textit{Precompute.}
The precompute component is responsible for the generation of the prior information that is jointly used with the refined training media $\mathcal{\hat{S}}$ in learning. For instance, global statistics and external resources can be used to synthesize new prior knowledge useful in learning.
The prior information commonly used is tag statistics in $\mathcal{S}$, including tag occurrence and tag co-occurrence.
Tag occurrence is used in \cite{tmm09-xirong} as a penalty to suppress overly frequent tags.
Measuring the semantic similarity between two tags is important for tag relevance learning algorithms that exploit tag correlations.
While linguistic metrics as those derived from WordNet were used before the proliferation of social media \cite{mm2005-yjin,mm2006-cwang},
they do not directly reflect how people tag images. 
For instance, tag `sunset' and tag `sea' are weakly related according to the WordNet ontology, 
but they often appear together in social tagging as many of the sunset photos are shot around seasides. 
Therefore, similarity measures that are based on tag statistics computed from many socially tagged images are in dominant use.
Sigurbj{\"o}rnsson and van Zwol utilized the Jaccard coefficient and a conditional tag probability in their tag suggestion system \cite{www2008-borkur},
while Liu \etal used normalized tag co-occurrence \cite{neurocom2013-jliu}.
To better capture the visual relationship between two tags, 
Wu \etal proposed the Flickr distance \cite{mm2008-lwu}.
The authors represent each tag by a visual language model, trained on bag of visual words features of images labeled with this tag.
The Flickr distance between two tags is computed as the Jensen-Shannon divergence between the corresponding models.
Later, Jiang \etal introduced the Flickr context similarity, which also captures the visual relationship between two tags, 
but without the need of the expensive visual modeling \cite{mm2009-yjiang}.
The trick is to compute the Normalized Google Distance \cite{tkde2004-rudi} between two tags,
but with tag statistics acquired from Flickr image collections instead of Google indexed web pages.
For its simplicity and effectiveness, 
we observe a prevalent use of the Flickr context similarity in the literature \cite{www09-liu,mm10-zhu,tmm10-wang,wsdm2011-zhuang,tmm12-zhu,tip13-gao,mm13-xirong,tcyb2014-qian}.

\section{A New Experimental Protocol} \label{sec:protocol}
In spite of the expanding literature,
there is a lack of consensus on the performance of the individual methods.
This is largely due to the fact that existing works either use homemade data, see \cite{www09-liu,tmm10-wang,tmm12-chen,tip13-gao}, which are not publicly accessible, or use selected subsets of benchmark data, e.g.~as in \cite{mm10-zhu,tmm12-sang,eccv2014-zfeng}. 
As a consequence, the performance scores reported in the literature are not comparable across the papers.

Benchmark data with manually verified labels is crucial for an objective evaluation. As Flickr has been well recognized as a profound manifestation of social image tagging, Flickr images act as a main source for benchmark construction. MIRFlickr from the Leiden University \cite{mir10-huiskes} and NUS-WIDE from the National University of Singapore \cite{civr09-chua} are the two most popular Flickr-based benchmark sets for social image tagging and retrieval, as demonstrated by the number of citations. On the use of the benchmarks, one typically follows a single-set protocol, that is, learning the underlying tag relevance function from the training part of a chosen benchmark set, and evaluating it on the test part. Such a protocol is inadequate given the dynamic nature of social media, which could easily make an existing benchmark set outdated.
For any method targeting at social images, a cross-set evaluation is necessary to test its generalization ability, which is however overlooked in the literature.

Another desirable property is the capability to learn from the increasing amounts of socially tagged images. Since existing works mostly use training data of a fixed scale, this property has not been well evaluated.

Following these considerations, we present a new experimental protocol, wherein training and test data from distinct research groups are chosen for evaluating a number of representative works in the cross-set scenario. Training sets with their size ranging from 10k to one million images are constructed to evaluate methods of varied complexity. To the best of our knowledge, such a comparison between many methods on varied scale datasets with a common experimental setup has not been conducted before. For the sake of experimental reproducibility, all data and code are  available online\footnoteref{codeurl}.

\subsection{Datasets} \label{ssec:exp-setup}

We describe the training media $\mathcal{S}$ and the test media $\mathcal{X}$ as follows,
with basic data characteristics and their usage summarized in Table \ref{tab:datasets}.

\emph{Training media} $\mathcal{S}$. We use a set of 1.2 million Flickr images collected by the University of Amsterdam~\cite{tmm12-li}, by using over 25,000 nouns in WordNet as queries to uniformly sample images uploaded between 2006 and 2010. 
Based on our observation that batch-tagged images, namely those labeled with the same tags by the same user, tend to be near duplicate, we have excluded these images beforehand. 
Other than this, we do not perform near-duplicate image removal.
To meet with methods that cannot handle large data, 
we created two random subsets from the entire training sets, 
resulting in three training sets of varied sizes, termed as Train10k, Train100k, and Train1m, respectively.

\emph{Test media}  $\mathcal{X}$. 
We use MIRFlickr \cite{mir10-huiskes} and NUS-WIDE \cite{civr09-chua} for tag assignment and refinement, as in \cite{mir2010-verbeek,mm10-zhu,icme13-uricchio} and \cite{tist11-tang,eccv2012-mcauley,mm10-zhu,icme13-uricchio} respectively. 
We use NUS-WIDE for evaluating tag retrieval as in \cite{jasist11-sun,iccv2011-wli}. In addition, for retrieval we collected another test set namely Flickr51 contributed by Microsoft Research Asia \cite{tmm10-wang,tip13-gao}.
The MIRFlickr set contains 25,000 images with ground truth available for 14 tags.
The NUS-WIDE set contains 259,233 images,
with ground truth available for 81 tags.
The Flickr51 set consists of 81,541 Flickr images with partial ground truth provided for 55 test tags.
Among the 55 tags, there are 4 tags which either have zero occurrence in our training data or have no correspondence in WordNet,
so we ignore them. 
Differently from the binary judgments in NUS-WIDE, 
Flickr51 provides graded relevance, with 0, 1, and 2 to indicate irrelevant, relevant, and very relevant, respectively.
Moreover, the set contains several ambiguous tags such as `apple' and `jaguar',
where relevant instances could exhibit completely different imagery, e.g., Apple computers versus fruit apples.
Following the original intention of the datasets, we use MIRFlickr and NUS-WIDE for evaluating tag assignment and tag refinement,
and Flickr51 and NUS-WIDE for tag retrieval.
For all the three test sets, we use the full dataset for testing.

Although the training and test media are all from Flickr, they were collected independently,
and consequently they have a relatively small amount of images overlapped with each other, as shown in Table \ref{tab:datasets2}.

\begin{table} [tb!]
\renewcommand{\arraystretch}{1.3}
\tbl{Our proposed experimental protocol instantiates the \emph{Media} and \emph{Tasks} dimensions of Fig. \ref{fig:pipeline} with three training sets and three test sets for tag assignment, refinement and retrieval. Note that the training sets are socially tagged, they have no ground truth available for any tag. \label{tab:datasets}
}
{\centering
\scalebox{0.8}{
\begin{tabular}{@{}l r r r r r r r r@{}}
\toprule

    & \multicolumn{4}{c}{\textbf{Media characteristics}} && \multicolumn{3}{c}{\textbf{Tasks}} \\
    \cmidrule{2-5} \cmidrule{7-9}
\textbf{Media} & \textbf{\# images} & \textbf{\# tags} & \textbf{\# users} & \textbf{\# test tags} && \textbf{assignment} & \textbf{refinement} & \textbf{retrieval}\\
\cmidrule{1-5} \cmidrule{7-9}
\emph{\textbf{Training media $\mathcal{S}$:}} \\
Train10k    & 10,000    &  41,253  & 9,249     & -- && \checked & \checked & \checked\\
Train100k   & 100,000   & 214,666  & 68,215    & -- && \checked & \checked & \checked\\
Train1m \cite{tmm12-li}     & 1,198,818  & 1,127,139 & 347,369  & -- && \checked & \checked & \checked\\ [3pt]
\emph{\textbf{Test media $\mathcal{X}$:}} \\
MIRFlickr \cite{mir10-huiskes}  & 25,000    &  67,389 &  9,862    & 14 && \checked & \checked & -- \\
Flickr51 \cite{tmm10-wang}      & 81,541    &  66,900 & 20,886    & 51 && -- & -- & \checked\\
NUS-WIDE \cite{civr09-chua}     & 259,233   & 355,913 & 51,645    & 81 && \checked & \checked & \checked \\

\bottomrule
\end{tabular}
}
}
\end{table}

\begin{table} [tb!]
\renewcommand{\arraystretch}{1.3}
\tbl{Data overlap between Train1M and the three test sets, 
measured in terms of the number of shared images, tags, and users, respectively.
Tag overlap is counted on the top 1,000 most frequent tags.
As the original photo ids of MIRFlickr have been anonymized, 
we cannot check image overlap between this dataset and Train1M.\label{tab:datasets2}
}
{\centering
\scalebox{0.8}{
\begin{tabular}{@{} p{5cm} l r r r p{5cm} @{}}
\toprule

\quad  &                 &  \multicolumn{3}{c}{\textbf{Overlap with Train1M}} & \quad\\
\cmidrule{3-5}
\quad & \textbf{Test media}  & \textbf{\# images} & \textbf{\# tags} & \textbf{\# users} & \quad\\
\cmidrule{2-5}
\quad & MIRFlickr    &  $-$    &  693  &   6,515 & \quad \\
\quad & Flickr51     &  730    &  538  &  14,211 & \quad \\
\quad & NUS-WIDE     &  7,975  &  718  &  38,481 & \quad \\

\bottomrule
\end{tabular}
}
}
\end{table}

\subsection{Implementation}
\label{ssec:implementation}

This section describes common implementations applicable to all the three tasks, including the choice of visual features and tag preprocessing.
Implementations that are applied uniquely to single tasks will be described in the coming sections.

\emph{Visual features}.
Two types of features are extracted to provide insights of the performance improvement achievable by appropriate feature selection: the classical bag of visual words (BoVW) and the current state of the art deep learning based features extracted from Convolutional Neural Networks (CNN).
The BoVW feature is extracted by the color descriptor software \cite{koen-tpami10}.
SIFT descriptors are computed at dense sampled points, at every 6 pixels for two scales.
A codebook of size 1,024 is created by K-means clustering.
The SIFTs are quantized by the codebook using hard assignment, and aggregated by sum pooling.
In addition, we extract a compact 64-d global feature \cite{color64-icme07},
combining a 44-d color correlogram, a 14-d texture moment, and a 6-d RGB color moment, to compensate the BoVW feature.
The CNN feature is extracted by the pre-trained VGGNet \cite{iclr2015-vggnet}. 
In particular, we adopt the 16-layer VGGNet, and take as feature vectors the last fully connected layer of ReLU activation,
resulting in a feature vector of 4,096 dimensions per image.
The BoVW feature is used with the $l_1$ distance and the CNN feature is used with the cosine distance for their good performance.

\textit{Vocabulary} $\mathcal{V}$. 
As what tags a person may use is meant to be open, the need of specifying a tag vocabulary is merely an engineering convenience.
For a tag to be meaningfully modeled, there has to be a reasonable amount of training images with respect to that tag.
For methods where tags are processed independently from the others, the size of the vocabulary has no impact on the performance.
In the other cases, in particular for transductive methods that rely on the image-tag association matrix, the tag dimension has to be constrained to make the methods runnable.
In our case, for these methods a three-step automatic cleaning procedure is performed on the training datasets. 
First, all the tags are lemmatized to their base forms by the NLTK software \cite{nltk}. 
Second, tags not defined in WordNet are removed. 
Finally, in order to avoid insufficient sampling, we remove tags that cannot meet a threshold on tag occurrence. 
The thresholds are empirically set as 50, 250, and 750 for Train10k, Train100k, and Train1m, respectively, 
in order to have a linear increase in vocabulary size versus a logarithmic increase in the number of labeled images. 
This results in a final vocabulary of 237, 419, and 1,549 tags, respectively, with all the test tags included.
Note that these numbers of tags are larger than the number of tags that can be actually evaluated.
This allows us to build a unified evaluation framework that is more handy for cross-dataset evaluation. 

\subsection{Evaluating tag assignment} \label{ssec:setup-assignment}

\textit{Evaluation criteria}. 
A good method for tag assignment shall rank relevant tags before irrelevant tags for a given test image.
Moreover, with the assigned tags, relevant images shall be ranked before irrelevant images for a given test tag.
We therefore use the image-centric Mean image Average Precision (MiAP) to measure the quality of tag ranking, 
and the tag-centric Mean Average Precision (MAP) to measure the quality of image ranking.
Let $m_{gt}$ be the number of ground-truthed test tags,
which is 14 for MIRFlickr and 81 for NUS-WIDE.
The image-centric Average Precision of a given test image $x$ is computed as
\begin{equation} \label{eq:miap}
     iAP(x) := \frac{1}{R} \sum_{j = 1}^{m_{gt}}\frac{r_j}{j} \delta(x,t_j),
  \end{equation}
where $R$ is the number of relevant tags of the given image,
$r_j$ is the number of relevant tags in the top $j$ ranked tags, and $\delta(x_i,t_j)= 1$ if tag $t_j$ is relevant and 0 otherwise. 
MiAP is obtained by averaging $iAP(x)$ over the test images.

The tag-centric Average Precision of a given test tag $t$ is computed as 
\begin{equation} \label{eq:MAP}
    AP(t) := \frac{1}{R} \sum_{i = 1}^{n} \frac{r_i}{i}\delta(x_i,t),
\end{equation}
where $R$ is the number of relevant images for the given tag, and $r_i$ is the number of relevant images in the top $i$ ranked images.
MAP is obtained by averaging $AP(t)$ over the test tags.

The two metrics are complementary to some extent. 
Since MiAP is averaged over images, each test image contributes equally to MiAP, as opposed to MAP where each tag contributes equally.
Consequently, MiAP is biased towards frequent tags,
while MAP can be easily affected by the performance of rare tags, especially when $m_{gt}$ is relatively small.
%

\textit{Baseline}. Any method targeting at tag assignment shall be better than a random guess,
which simply returns a random set of tags.
The RandomGuess baseline is obtained by computing MiAP and MAP given the random prediction, 
which is run 100 times with the resulting scores averaged.

\subsection{Evaluating tag refinement} \label{ssec:setup-refinement}
\textit{Evaluation criteria}. As tag refinement is also meant for improving tag ranking and image ranking,
it is evaluated by the same criteria, i.e., MiAP and MAP, as used for tag assignment.

\textit{Baseline}. 
A natural baseline for tag refinement is the original user tags assigned to an image, which we term as UserTags.

\subsection{Evaluating tag retrieval} \label{ssec:setup-retrieval}
\textit{Evaluation criteria}. 
To compare methods for tag retrieval, for each test tag we first conduct tag-based image search to retrieve images labeled with that tag, 
and then sort the images by the tag relevance scores. 
We use MAP to measure the quality of the entire image ranking.
As users often look at the top ranked results and hardly go through the entire list, we also report Normalized Discounted Cumulative Gain (NDCG), commonly used to evaluate the top few ranked results of an information retrieval system \cite{ndcg}.
Given a test tag $t$, its NDCG at a particular rank position $h$ is defined as:
\begin{equation}
NDCG_h(t) :=  \frac{DCG_h(t)}{IDCG_h(t)},
\end{equation}
where 
$DCG_h(t) = \sum_{i=1}^h \frac{2^{rel_i}-1}{\log_2(i+1)}$,
$rel_i$ is the graded relevance of the result at position $i$,
and $IDCG_h$ is the maximum possible $DCG$ till position $h$.
We set $h$ to be 20, which corresponds to a typical number of search results presented on the first two pages of a web search engine.
Similar to MAP, NDCG$_{20}$ of a specific method on a specific test set is averaged over the test tags of that test set.

\textit{Baselines}. 
When searching for relevant images for a given tag, it is natural to ask how much a specific method gains compared to a baseline system which simply returns a random subset of images labeled with that tag. 
Similar to the refinement baseline, we also denote this baseline as UserTags, as both of them purely use the original user tags.
For each test tag, the test images labeled with this tag are sorted at random, and MAP and NDCG$_{20}$ are computed accordingly.
The process is executed 100 times, and the average score over the 100 runs is reported.

The number of tags per image is often included for image ranking in previous works \cite{www09-liu,mm09-hxu}. 
Hence, we build another baseline system, denoted as TagNum, which sort images in ascending order by the number of tags per image.
The third baseline, denoted as TagPosition, is from \cite{jasist11-sun}, 
where the relevance score of a tag is determined by its position in the original tag list uploaded by the user.
More precisely, the score is computed as $1-position(t)/l$, where $l$ is the number of tags.

\section{Methods Selected for Comparison} \label{sec:selected}
Despite the rich literature, most works do not provide code. An exhaustive evaluation covering all published methods is impractical. We have to leave out methods that do not show significant improvements or novelties w.r.t. the seminal papers in the field, and methods that are difficult to replicate with the same mathematical preciseness as intended by their developers. 
We drive our choice by the intention to cover methods that aim for each of the three tasks, exploiting varied modalities by distinct learning mechanisms.
Eventually we evaluate 11 representative methods. For each method we analyze its scalability in terms of both computation and memory. Our analysis leaves out operations that are independent of specific tags and thus only need to be executed once in an offline manner, such as visual feature extraction, tag preprocessing, prior information precomputing, and filtering. Main properties of the methods are summarized in table \ref{tab:methods}.
Concerning the choices of parameters, we adopt what the original papers recommend. 
When no recommendation is given for a specific method, we try a range of values to our best understanding,
and choose the parameters that yield the best overall performance.

\subsection{Methods under analysis}

\textbf{1. SemanticField} \cite{tmm12-zhu}.
This method measures tag relevance in terms of an averaged semantic similarity between the tag and the other tags assigned to the image:
\begin{equation} \label{eq:semfield}
f_{SemField}(x,t) := \frac{1}{l_x} \sum_{i=1}^{l_x} sim(t,t_i),
\end{equation}
where $\{t_1,\ldots,t_{l_x}\}$ is a list of $l_x$ social tags assigned to the image $x$,
and $sim(t,t_i)$ denotes a semantic similarity between two tags. 
SemanticField explicitly assumes that several tags are associated to visual data and their coexistence is accounted in the evaluation of tag relevance.
Following \cite{tmm12-zhu},
the similarity is computed by combining the Flickr context similarity and the WordNet Wu-Palmer similarity \cite{acl94-wup}. 
The WordNet based similarity exploits path length in the WordNet hierarchy to infer tag relatedness.
We make a small revision of \cite{tmm12-zhu}, i.e.~combining the two similarities by averaging instead of multiplication, 
because the former strategy produces slightly better results.
SemanticField requires no training except for computing tag-wise similarity, which can be computed offline and is thus omitted. 
Having all tag-wise similarities in memory, applying Eq. (\ref{eq:semfield}) requires $l_x$ table lookups per tag. 
Hence, the computational complexity is $O(m\cdot l_x)$, and $O(m^2)$ for memory.
\smallskip

\textbf{2. TagRanking} \cite{www09-liu}.
The tag ranking algorithm consists of two steps. Given an image $x$ and its tags, the first step produces an initial tag relevance score for each of the tags, obtained by (Gaussian) kernel density estimation on a set of $\bar{n}=1,000$ images labeled with each tag, separately.
Secondly, a random walk is performed on a tag graph where the edges are weighted by a tag-wise similarity. 
We use the same similarity as in SemanticField.
Notice that when applied for tag retrieval, the algorithm uses the rank of $t$ instead of its score, i.e.,
\begin{equation} \label{eq:tagranking}
f_{TagRanking}(x,t) = -rank(t) + \frac{1}{l_x},
\end{equation}
where $rank(t)$ returns the rank of $t$ produced by the tag ranking algorithm. 
The term $\frac{1}{l_x}$  is a tie-breaker when two images have the same tag rank.
Hence, for a given tag $t$, TagRanking cannot distinguish relevant images from irrelevant images if $t$ is the sole tag assigned to them. It explicitly exploits the coexistence of several tags per image.
TagRanking has no learning stage. To derive tag ranks for Eq. \ref{eq:tagranking}, 
the main computation is the kernel density estimation on $\bar{n}$ socially-tagged examples for each tag,
followed by an $L$ iteration random walk on the tag graph of $m$ nodes.
All this results in a computation cost of $O(m \cdot d \cdot \bar{n} + L \cdot {m}^2)$ per test image.
Because the two steps are executed sequentially, the corresponding memory cost is $O(\max(d \bar{n}, m^2))$.
\smallskip

\textbf{3. KNN} \cite{ijcv10-makadia}.
This algorithm estimates the relevance of a given tag with respect to an image by 
first retrieving $k$ nearest neighbors from $\mathcal{S}$ based on a visual distance $d$, 
and then counting the tag occurrence in associated tags of the neighborhood. 
In particular, KNN builds $f_\Phi(x,t;\Theta)$ as:
\begin{equation} \label{eq:KNN}
f_{KNN}(x,t) := k_t,
\end{equation}
where $k_t$ is the number of images with $t$ in the visual neighborhood of $x$.
The instance-based KNN requires no training. 
The main computation of $f_{KNN}$ is to find $k$ nearest neighbors from $\mathcal{S}$, 
which has a complexity of $O(d \cdot |\mathcal{S}| + k \cdot \log |\mathcal{S}| )$ per test image, 
and a memory footprint of $O(d \cdot |\mathcal{S}|)$ to store all the $d$-dimensional feature vectors.
It is worth noting that these complexities are drawn from a straightforward implementation of $k$-nn search, 
and can be substantially reduced by employing more efficient search techniques, c.f. \cite{pami2011-jegou}.
Accelerating KNN by the product quantization technique \cite{pami2011-jegou} imposes an extra training step, 
where one has to construct multiple vector quantizers by K-means clustering, and further use the quantizers to compress the original feature vector into a few codes.
\smallskip

\textbf{4. TagVote} \cite{tmm09-xirong}. 
The TagVote algorithm estimates the relevance of a tag $t$ w.r.t.~an image $x$ by counting the occurrence frequency of $t$ in social annotations of the visual neighbors of $x$. 
 Different from KNN, TagVote exploits the user element, introducing a unique-user constraint on the neighbor set to make the voting result more objective.
 Each user has at most one image in the neighbor set. 
 Moreover, TagVote takes into account tag prior frequency to suppress over frequent tags.
In particular, the TagVote algorithm builds $f_\Phi(x,t;\Theta)$ as 
\begin{equation} \label{eq:tagvote}
f_{TagVote}(x, t) := k_t - k \frac{n_t}{|\mathcal{S}|},
\end{equation}
where $n_t$ is the number of images labeled with $t$ in $\mathcal{S}$. 
Following \cite{tmm09-xirong}, we set $k$ to be 1,000 for both KNN and TagVote.
TagVote has the same order of complexity as KNN.
\smallskip

\textbf{5. TagProp} \cite{iccv09-tagprop,mir2010-verbeek}.
TagProp employs neighbor voting plus distance metric learning. 
A probabilistic framework is proposed where the probability of using images in the neighborhood is defined based on rank or distance-based weights.
TagProp builds $f_\Phi(x,t;\Theta)$ as:
\begin{equation} \label{eq:tagprop}
f_{TagProp}(x, t) := \sum_j^k  \pi_j \cdot \mathbf{I}(x_j,t), 
\end{equation}
where $\pi_j$ is a non-negative weight indicating the importance of the $j$-th neighbor $x_j$,
and $\mathbf{I}(x_j,t)$ returns 1 if $x_j$ is labeled with $t$, and 0 otherwise.
Following \cite{mir2010-verbeek}, we use $k=1,000$ and the rank-based weights,
which showed similar performance to the distance-based weights.
Different from TagVote that uses tag prior to penalize frequent tags, TagProp promotes rare tags and penalizes frequent ones by training a logistic model per tag upon $f_{TagProp}(x, t)$.
The use of the logistic model makes TagProp a model-based method.
In contrast to KNN and TagVote wherein visual neighbors are treated equally, 
TagProp employs distance metric learning to re-weight the neighbors, yielding a learning complexity of $O(l \cdot m \cdot k)$ where $l$ is the number of gradient descent iterations it needs (typically less than 10).
TagProp maintains $2m$ extra parameters for the logistic models, though their storage cost is ignorable compared to the visual features.
Therefore, running Eq. (\ref{eq:tagprop}) has the same order of complexity as KNN and TagVote.
\smallskip

\textbf{6. TagCooccur} \cite{www2008-borkur}.
While both SemanticField and TagCooccur are tag-based,
the main difference lies in how they compute the contribution of a specific tag to the test tag's relevance score.
Different from SemanticField which uses tag similarities,
TagCooccur uses the test tag's rank in the tag ranking list created by sorting all tags in terms of their co-occurrence frequency with the tag in $\mathcal{S}$.
In addition, TagCooccur takes into account the stability of the tag, measured by its frequency.
The method is implemented as
\begin{equation} \label{eq:tagco}
f_{tagcooccur}(x,t)= descriptive(t) \sum_{i=1}^{l_x} vote(t_i,t) \cdot  rank\mbox{-}promotion(t_i,t) \cdot stability(t_i),
\end{equation}
where $descriptive(t)$ is to damp the contribution of tags with a very high-frequency,
$rank\mbox{-}promotion(t_i,t)$ measures the rank-based contribution of $t_i$ to $t$, 
$stability(t_i)$ for promoting tags for which the statistics are more stable,
and $vote(t_i,t)$ is 1 if $t$ is among the top 25 ranked tags of $t_i$, and 0 otherwise.
TagCooccur has the same order of complexity as SemanticField.

\textbf{7. TagCooccur+} \cite{tmm09-xirong}.
TagCooccur+ is proposed to improve TagCooccur by adding the visual content.
This is achieved by multiplying $f_{tagcooccur}(x,t)$ with a content-based term, i.e.,
\begin{equation} \label{eq:tagcoplus}
f_{tagcooccur+}(x,t) = f_{tagcooccur}(x,t) \cdot \frac{k_c}{k_c + r_c(t) - 1},
\end{equation}
where $r_c(t)$ is the rank of $t$ when sorting the vocabulary by $f_{TagVote}(x, t)$ in descending order,
and $k_c$ is a positive weighting parameter, which is empirically set to 1.
While TagCooccur+ is grounded on TagCooccur and TagVote, the complexity of the former is ignorable compared to the latter,
so the complexity of TagCooccurs+ is the same as KNN.

\textbf{8. TagFeature} \cite{tmm12-chen}. 
The basic idea is to enrich image features by adding an extra tag feature. 
A tag vocabulary that consists of $d'$ most frequent tags in $\mathcal{S}$ is constructed first.
Then, for each tag a two-class linear SVM classifier is trained using LIBLINEAR \cite{jmlr2008-liblinear}.
The positive training set consists of $p$ images labeled with the tag in $\mathcal{S}$, 
and the same amount of negative training examples are randomly sampled from images not labeled with the tag.
The probabilistic output of the classifier, obtained by the Platt's scaling \cite{note-platt-jml07}, corresponds to a specific dimension in the tag feature.
By concatenating the tag and visual features,
an augmented feature of $d + d'$ dimension is obtained. 
For a test tag $t$, its tag relevance function $f_{TagFeature}(x,t)$ is obtained by re-training an SVM classifier using the augmented feature.
The linear property of the classifier allows us to first sum up all the support vectors into a single vector 
and consequently to classify a test image by the inner product with this vector. That is,
\begin{equation} \label{eq:svm-af}
f_{TagFeature}(x,t) := b + <x_t, x>,
\end{equation}
where $x_t$ is the weighted sum of all support vectors and $b$ the intercept.
To build meaningful classifiers, we use tags that have at least 100 positive examples.
While $d'$ is chosen to be 400 in \cite{tmm12-chen},
the two smaller training sets, namely Train10k and Train100k, have 76 and 396 tags satisfying the above requirement.
We empirically set $p$ to 500, and do random down-sampling if the amount of images for a tag exceeds this number.
For TagFeature, learning a linear classifier for each tag from $p$ positive and $p$ negative examples requires $O((d+d')p)$ in computation and $O((d+d')p)$ 
in memory \cite{jmlr2008-liblinear}.
Running Eq. (\ref{eq:svm-af}) for all the $m$ tags and $n$ images needs $O(n m(d+d'))$ in computation and $O(m(d+d'))$ in memory.

\smallskip

\textbf{9. RelExample} \cite{mm13-xirong}. 
Different from TagFeature \cite{tmm12-chen} that directly learns from tagged images,
RelExample exploits positive and negative training examples which are deemed to be more relevant with respect to the test tag $t$. 
In particular, relevant positive examples are selected from $\mathcal{S}$ by combining SemanticField and TagVote in a late fusion manner.   
For negative training example acquisition, they leverage Negative Bootstrap \cite{tmm2013-xli}, 
a negative sampling algorithm which iteratively selects negative examples deemed most relevant for improving classification. 
A $T$-iteration Negative Bootstrap will produce $T$ meta classifiers.
The corresponding tag relevance function is written as 
\begin{equation} \label{eq:svm-re}
f_{RelExample}(x,t) := \frac{1}{T} \sum_{l=1}^T (b_l + \sum_{j=1}^{n_l} \alpha_{l,j} \cdot y_{l,j} \cdot \mathcal{K}(x,x_{l,j})),
\end{equation}
where  
$\alpha_{l,j}$ is a positive coefficient of support vector $x_{l,j}$, 
$y_{l,j} \in \{-1,1\}$ is class label,
and $n_l$ the number of support vectors in the $l$-th classifier.
For the sake of efficiency, the kernel function $\mathcal{K}$ is instantiated with the fast intersection kernel \cite{cvpr2008-smaji}.
RelExample uses the same amount of positive training examples as TagFeature.
The number of iterations $T$ is empirically set to 10.
For the SVM classifiers used in TagFeature and RelExample,
the Platt's scaling \cite{note-platt-jml07} is employed to convert prediction scores into probabilistic output. 
In RelExample, for each tag learning a histogram intersection kernel SVM has a computation cost of $O(dp^2)$ per iteration,
and $O(T dp^2)$ for $T$ iterations. 
By jointly using the fast intersection kernel with a quantization factor of $q$ \cite{cvpr2008-smaji} and model compression \cite{tmm2013-xli}, 
an order of $O(dq)$ is needed to keep all learned meta classifiers in memory.
Since learning a new classifier needs a memory of $O(dp)$, the overall memory cost for training RelExample is $O(dp + dq)$.
For each tag, model compression is applied to its learned ensemble in advance to running Eq. (\ref{eq:svm-re}).
As a consequence, the compressed classifier can be cached in an order of $O(dq)$ and executed in an order of $O(d)$.
\smallskip

\textbf{10. RobustPCA} \cite{mm10-zhu}. 
On the base of robust principal component analysis \cite{jacm2011-candes}, 
RobustPCA factorizes the image-tag matrix $D$ by a low rank decomposition with error sparsity. 
That is,
\begin{equation} \label{eq:lres-add}
    D = \hat{D} + E,
\end{equation}
where the reconstructed $\hat{D}$ has a low rank constraint based on the nuclear norm,
and $E$ is an error matrix with a $\ell_1$-norm sparsity constraint. 
Notice that the decomposition is not unique. 
So for a better solution, the decomposition process takes into account image affinities and tag affinities, 
by adding two extra penalties with respect to a Laplacian matrix $L_i$ from the image affinity graph 
and another Laplacian matrix $L_t$ from the tag affinity graph.
Consequently, two hyper-parameters $\lambda_1$ and $\lambda_2$ 
are introduced to balance the error sparsity and the two Laplacian strengths. 
We follow the original paper and set the two parameters by performing a grid search on the very same proposed range. 
To address the tag sparseness, the authors employ a preprocessing step to refine $D$ by a weighted KNN propagation based on the visual similarity.
RobustPCA requires an iterative procedure based on the accelerated proximal gradient method with a quadratic convergence rate \cite{mm10-zhu}. 
Each iteration spends the majority of the required time performing Singular Value Decomposition that, according to \cite{golub2012}, has a well known complexity of $O(c m^2 n + c' n^3)$ where $c,c'$ are constants. 
Regarding memory, it has a requirement of $O(c n \cdot m + c' \cdot(n^2 + m^2))$ as it needs to maintain a full copy of $D$ and Laplacians of images and labels.
\smallskip

\textbf{11. TensorAnalysis} \cite{tmm12-sang}. 
This method considers ternary relationships between images, tags and user, by extending the image-tag association matrix 
to a binary user-image-tag tensor $F \in \{0, 1\}^{|\mathcal{X}| \times |\mathcal{V}| \times |\mathcal{U}|}$.
The tensor is factorized by Tucker decomposition into a dense core $C$ and three low rank matrices $U$, $I$, $T$, corresponding to the user, image, and tag modalities, respectively:
\begin{equation} \label{eq:rmtf}
    F = C \times_u U \times_i I \times_t T,
\end{equation}
Here $\times_j$ is the tensor product between a tensor and a matrix along dimension $j \in \{u,i,t\}$. 
The idea is that $C$ contains the interactions between modalities, while each low-rank matrix represents the main components of each modality. Every modality has to be sized manually or by energy retention, adding three needed parameters $R = (r_I, r_T, r_U)$. 
The tag relevance scores are obtained by computing $\hat{D} = C \times_i I \times_t T \times_u \mathbf{1}_{r_u}$.
Similar to RobustPCA, the decomposition in Eq. (\ref{eq:rmtf}) is not unique and a better solution may be found by regularizing the optimization process with a Laplacian built on a similarity graph for each modality, i.e., $L_i$, $L_t$, and $L_u$, and a 
$\ell_2$ regularizer on each factor i.e. $C$, $U$, $I$ and $T$.
For TensorAnalysis, the complexity is $O(|P_1| \cdot (r_T \cdot m^2 +r_U \cdot r_I \cdot r_T))$, proportional to the number of tags $P_1$ asserted in $D$ and the dimension of low rank $r_U,r_I,r_T$ factors. 
The memory required is $O(n^2 + m^2 + u^2)$ for the Laplacians of images, tags and users. 

\begin{table} [tb!]
\renewcommand{\arraystretch}{1.3}
\tbl{Main properties of the eleven methods evaluated in this survey following the dimensions of Fig. \ref{fig:pipeline}.
The computational and memory complexity of each method is based on processing $n$ test images and $m$ test tags by exploiting the training set $\mathcal{S}$. \label{tab:methods} }
{
\centering
\scalebox{0.65}{
\begin{tabular}{@{}l l l l ll l lllll@{}}
\toprule
 & &  &  & \multicolumn{2}{c}{\textbf{Auxiliary Component}}  &  & \multicolumn{5}{c}{\textbf{Learning}} \\ 
 
 \cmidrule{5-6} \cmidrule{8-12}

\textbf{Methods} & \textbf{Test Media} & \textbf{Task} && \emph{Filter} & \emph{Precompute}  &  & \emph{Train Computation} & \emph{Test Computation} & & \emph{Train Memory} & \emph{Test Memory}   \\ 

\cmidrule{1-12}

\emph{\textbf{Instance-based}:} \\

SemanticField & tag  &  Retrieval && WordNet   & $sim(t,t')$  & & -- & $O(n  m l_x)$ & & -- & $O(m^2)$\\
\cmidrule{2-12}

TagCooccur   & tag   & \specialcell{Refinement\\Retrieval} && -- & \specialcell{Tag prior\\Co-occurrence} &  & -- & $O(n  m l_x)$ & & -- & $O(m^2)$ \\
\cmidrule{2-12}

TagRanking  & tag + image  & Retrieval && -- & $sim(t,t')$  &   & -- & $O(n(m  d \bar{n} + L  m^2) )$ &  & -- & $O(\max(d \bar{n}, m^2))$\\
\cmidrule{2-12}

KNN   & tag + image & \specialcell{Assignment\\Retrieval}  && -- & --  &   & -- & $O(n(d |\mathcal{S}| + k \log |\mathcal{S}|) )$ & & -- & $O(d  |\mathcal{S}|)$\\
\cmidrule{2-12}

TagVote   & tag + image  & \specialcell{Assignment\\Retrieval} && -- & Tag prior &  & -- & $O(n(d  |\mathcal{S}| + k \log |\mathcal{S}|) )$ & & -- & $O(d  |\mathcal{S}|)$\\
\cmidrule{2-12}

TagCooccur+   & tag + image  & \specialcell{Refinement\\Retrieval} && -- & \specialcell{Tag prior\\Co-occurrence}  &  & -- & $O(n(d  |\mathcal{S}| + k \log |\mathcal{S}|) )$ & & -- & $O(d  |\mathcal{S}|)$\\

\cmidrule{1-12}

\emph{\textbf{Model-based}:} \\

TagProp   & tag + image & \specialcell{Assignment\\Retrieval} && -- & -- &  & $O(l \cdot m \cdot k)$  & $O(n(d |\mathcal{S}| + k  \log |\mathcal{S}|) )$  & & $O(d  |\mathcal{S}| + 2m)$ & $O(d  |\mathcal{S}| + 2m)$\\
\cmidrule{2-12}

TagFeature  & tag + image & \specialcell{Assignment\\Retrieval}  &&  -- & Tag classifiers  &   & $O(m (d+d') p)$ & $O(n m (d+d'))$ & &  $O( (d+d')p)$  & $O(m (d+d'))$\\
\cmidrule{2-12}

RelExample &  tag + image  & \specialcell{Assignment \\ Retrieval} && \specialcell{SemField \\ + TagVote}  &  $sim(t,t')$ & & $O(m T d p^2 )$ & $O(dp + dq)$ & & $O(n m d)$ & $O(m dq)$\\ [3pt]
\cmidrule{1-12}

 


\emph{\textbf{Transduction-based}:} \\

RobustPCA   & tag + image   & \specialcell{Refinement\\ Retrieval} && \specialcell{WordNet\\ + KNN}  & $L_i, L_t$      &   & \multicolumn{2}{c}{$O(c m^2 n + c' n^3)$} & & \multicolumn{2}{c}{$O(c n m + c' \cdot(n^2 + m^2))$}\\
\cmidrule{2-12}

TensorAnalysis  & tag + image + user   & Refinement && Postag sets & $L_i, L_t, L_u$  &  & \multicolumn{2}{c}{$O(|P_1| \cdot (r_T \cdot m^2 +r_U \cdot r_I \cdot r_T))$} && \multicolumn{2}{c}{$O(n^2 + m^2 + u^2)$}\\

\bottomrule
\end{tabular}
}}
\end{table}

\subsection{Considerations} \label{ssec:scalablity-analysis}
An overview of the methods analyzed is given Table \ref{tab:methods}.
Among them, SemanticField, counting solely on the tag modality, has the best scalability with respect to both computation and memory. 
Among the instance-based methods, TagRanking, which works on selected subsets of $\mathcal{S}$ rather than the entire collection, has the lowest memory request. When the number of tags to be modeled is substantially smaller than the size of $\mathcal{S}$, the model-based methods require less memory and run faster in the test stage, but at the expense of SVM model learning in the training stage. The two transduction-based methods have limited scalability, and can operate only on small sized $\mathcal{S}$.

\section{Evaluation} \label{sec:eval}


This section presents our evaluation of the 11 methods according to their applicability to the three tasks using the proposed experimental protocol, that is, KNN, TagVote, TagProp, TagFeature and RelExample for tag assignment (Section \ref{ssec:exp-tagassign}),
TagCooccur, TagCooccur+, RobustPCA, and TensorAnalysis for tag refinement (Section \ref{ssec:exp-tagrefine}),
and all for tag retrieval (Section \ref{ssec:exp-retrieval}). 
For TensorAnalysis we were able to evaluate only tag refinement with BovW features on MIRFlickr with Train10k and Train100k.  
The reason for this exception is that our implementation of TensorAnalysis performs worse than the baseline.
Consequently, the results of TensorAnalysis were kindly provided by the authors in the form of tag ranks.
Since the provided tag ranks cannot be converted to image ranks, we could not compute MAP scores.
A comparison between our Flickr based training data and ImageNet is given in Section \ref{ssec:imagenet}.


\subsection{Tag assignment} \label{ssec:exp-tagassign}

\begin{table} [tb!]
\renewcommand{\arraystretch}{1.2}
\tbl{Evaluating methods for tag assignment. Given the same feature, bold values indicate top performers on individual test sets. \label{tab:exp-autotag}
}
{\centering
\scalebox{1}{
\begin{tabular}{@{}l r r r r r r r@{}}
\toprule
   & \multicolumn{3}{c}{\textbf{MIRFlickr}} && \multicolumn{3}{c}{\textbf{NUS-WIDE}}  \\
                \cmidrule(lr){2-4} \cmidrule(l){6-8} 
 \textbf{Method} & Train10k &  Train100k & Train1m & & Train10k &  Train100k & Train1m \\
\cmidrule{1-8} 
\textit{\textbf{MiAP scores:}} \\
\rr RandomGuess       & 0.147 & 0.147 & 0.147           && 0.061 & 0.061 & 0.061  \\ [3pt]
\rr BovW + KNN        & 0.232 & 0.286 & 0.312           && 0.171 & 0.217 & 0.248 \\
\rr BovW + TagVote    & 0.276 & 0.310 & \textbf{0.328}  && 0.183 & 0.231 & 0.259  \\
\rr BovW + TagProp    & 0.276 & 0.299 & 0.314           && 0.230 & 0.249 & \textbf{0.268}  \\
\rr BovW + TagFeature & 0.278 & 0.294 & 0.298           && 0.244 & 0.221 & 0.214 \\
\rr BovW + RelExample & 0.284 & 0.309 & 0.303           && 0.257 & 0.233 & 0.245  \\ [3pt]

\rr CNN + KNN         & 0.326  & 0.366 & 0.379          && 0.315 & 0.343 & 0.376 \\
\rr CNN + TagVote     & 0.355  & 0.378 & 0.389          && 0.340 & 0.370 & \textbf{0.396} \\
\rr CNN + TagProp     & 0.373  & 0.384 & \textbf{0.392} && 0.366 & 0.376 & 0.380 \\ 
\rr CNN + TagFeature  & 0.359  & 0.378 & 0.383          && 0.367 & 0.338 & 0.373 \\ 
\rr CNN + RelExample  & 0.309  & 0.385 & 0.373          && 0.365 & 0.354 & 0.388 \\ [3pt]

\textit{\textbf{MAP scores:}} \\
\rr RandomGuess       & 0.072 & 0.072 & 0.072           && 0.023 & 0.023 & 0.023 \\ [3pt]
\rr BovW + KNN        & 0.231 & 0.282 & 0.336           && 0.094 & 0.139 & 0.185 \\
\rr BovW + TagVote    & 0.228 & 0.280 & 0.334           && 0.093 & 0.137 & 0.184  \\ 
\rr BovW + TagProp    & 0.245 & 0.293 & \textbf{0.342}  && 0.102 & 0.149 & \textbf{0.193} \\ 
\rr BovW + TagFeature & 0.200 & 0.199 & 0.201           && 0.090 & 0.096 & 0.098 \\
\rr BovW + RelExample & 0.284 & 0.303 & 0.310           && 0.119 & 0.155 & 0.172 \\  [3pt]

\rr CNN + KNN         & 0.564 & 0.613 & 0.639           && 0.271 & 0.356 & 0.400 \\
\rr CNN + TagVote     & 0.561 & 0.613 & 0.638           && 0.257 & 0.358 & \textbf{0.402} \\
\rr CNN + TagProp     & 0.586 & 0.619 & \textbf{0.641}  && 0.305 & 0.376 & 0.397\\
\rr CNN + TagFeature  & 0.444 & 0.554 & 0.563           && 0.262 & 0.310 & 0.326 \\
\rr CNN + RelExample  & 0.538 & 0.603 & 0.584           && 0.300 & 0.346 & 0.373 \\

\bottomrule
\end{tabular}
}}
\end{table}


Table \ref{tab:exp-autotag} shows the tag assignment performance of KNN, TagVote, TagProp, TagFeature and RelExample.
Their superior performance against the RandomGuess baseline shows that learning purely from social media is meaningful.
TagVote and TagProp are the two best performing methods on both test sets.
Substituting CNN for BovW consistently brings improvements for all methods.

\begin{figure}[!hbt]
\centering
                \includegraphics[width=1.\textwidth]{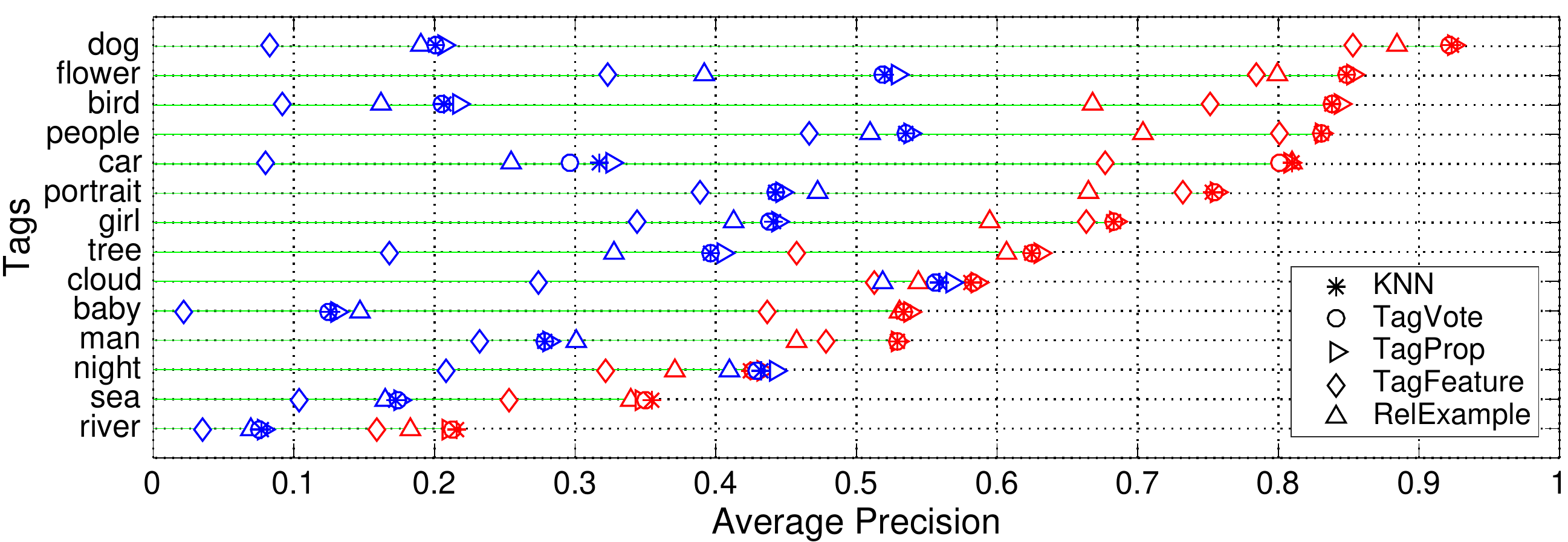}
\caption{
\textbf{Per-tag comparison of methods for tag assignment on MIRFlickr}, trained on Train1m.  The colors identify the features used: \textbf{blue} for BovW, \textbf{red} for CNN. The test tags have been sorted in descending order by the performance of CNN + TagProp.
} 
\label{fig:tag_comparison_mirflickr}
\end{figure}

In more detail, the following considerations hold. 
TagProp has higher MAP performance than KNN and TagVote in almost all the cases under analysis. 
As discussed in Section \ref{sec:selected}, 
TagProp is built upon KNN, but it weights the neighbor images by rank and applies a logistic model per tag. Since the logistic model does not affect the image ranking, the superior performance of TagProp should be ascribed to rank-based neighbor weighting.
A per-tag comparison on MIRFlickr is given in Fig. \ref{fig:tag_comparison_mirflickr}. TagProp is almost always ahead of KNN and TagVote.
Concerning TagVote and KNN, recall that their main difference is that TagVote applies the unique-user constraint on the neighborhood and it employs tag prior as a penalty term.
The fact that the training data contains no batch-tagged images minimizes the influence of the unique-user constraint.
While the penalty term does not affect image ranking for a given tag, it affects tag ranking for a given image. 
This explains why KNN and TagVote have mostly the same MAP.
Also, the result suggests that the tag prior based penalty is helpful for doing tag assignment by neighbor voting.

We observe that RelExample has a better MAP than TagFeature in every case. The absence of a filtering component makes TagFeature more likely to overfit to training examples irrelevant to the test tags. 
For the other two model-based methods, the overfit issue is alleviated by different strategies: RelExample employs a filtering component to select more relevant training examples, while TagProp has less parameters to tune.

\begin{figure}[!hbt]
\centering
                \includegraphics[width=1.\textwidth]{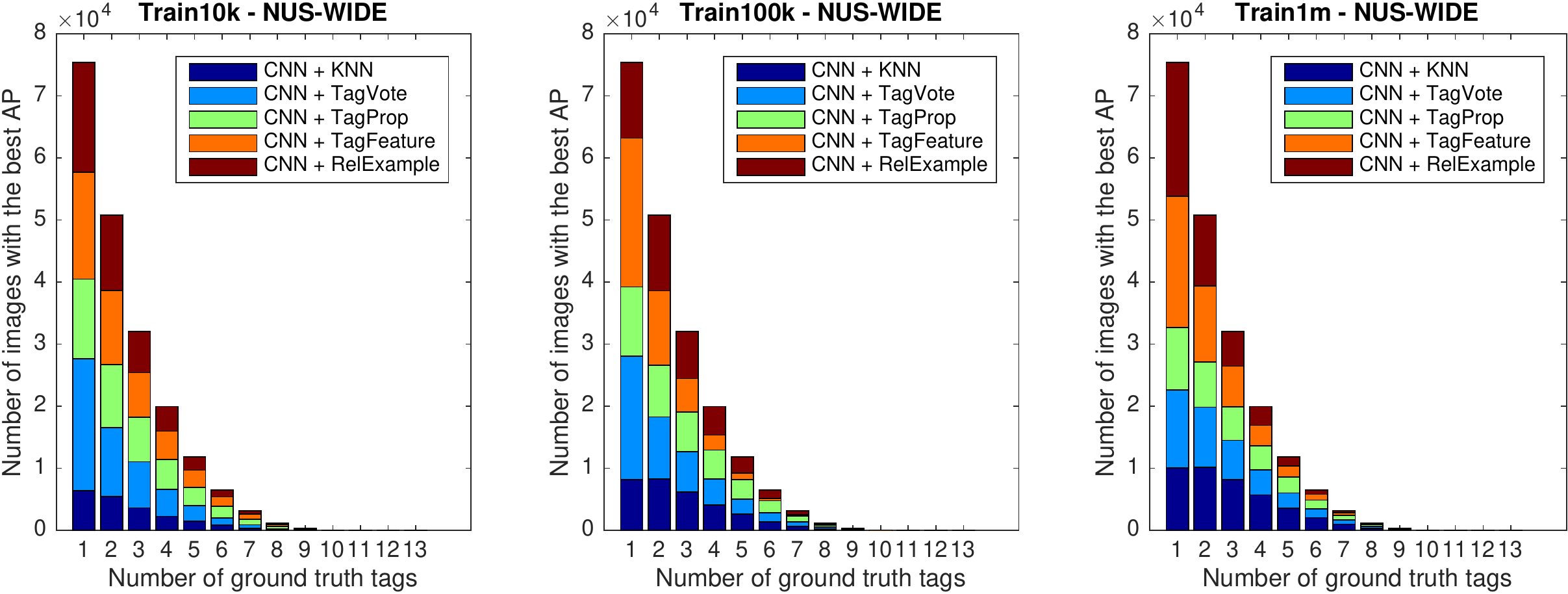}
\caption{
\textbf{Per-image comparison of methods for tag assignment on NUS-WIDE}. Test images are grouped in terms of their number of ground truth tags. The area of a colored bar is proportional to the number of images that the corresponding method scores best.
} 
\label{fig:miap_comparison_nuswide}
\end{figure}

A per-image comparison on NUS-WIDE is given in Fig. \ref{fig:miap_comparison_nuswide}.
The test images are put into disjoint groups so that images within the same group have the same number of ground truth tags.
For each group, the area of the colored bars is proportional to the number of images on which the corresponding methods score best.
The first group, i.e., images containing only one ground-truth tag, has the most noticeable change as the training set grows.
There are 75,378 images in this group, and for 39\% of the images, their single label is `person'.  
When Train1m is used,  RelExample beats KNN, TagVote, and TagProp for this frequent label. 
This explains the leading position of RelExample in the first group. 
The result also confirms our earlier discussion in Section \ref{ssec:setup-assignment} that MiAP is likely to be biased by frequent tags.

In summary, as long as enough training examples are provided, instance-based methods are on par with model-based methods for tag assignment.
Model-based methods are more suited when the training data is of limited availability.
However, they are less resilient to noise, and consequently a proper filtering strategy for refining the training data becomes essential.

\subsection{Tag refinement} \label{ssec:exp-tagrefine}

Table \ref{tab:exp-tagrefine} shows the performance of different methods for tag refinement.
We were unable to complete the table. In particular, RobustPCA could not go over 350k images due to its high demand in both CPU time and memory (see Table \ref{tab:methods}), while TensorAnalysis was provided by the authors only on MIRFlickr with Train10k, Train100k, and the BovW feature. 

\begin{table} [tb!]
\renewcommand{\arraystretch}{1.3}
\tbl{Evaluating methods for tag refinement. The asterisk (*) indicates results provided by the authors of the corresponding methods,
while the dash (--) means we were unable to produce results.
Given the same feature, bold values indicate top performers on individual test sets per performance metric.
\label{tab:exp-tagrefine}}
{
\centering
\scalebox{1}{
\begin{tabular}{@{}l r r r r r r r@{}}
\toprule
    & \multicolumn{3}{c}{\textbf{MIRFlickr}} && \multicolumn{3}{c}{\textbf{NUS-WIDE}}  \\
                \cmidrule(lr){2-4} \cmidrule(l){6-8} 
\textbf{Method} & Train10k &  Train100k & Train1m & & Train10k &  Train100k & Train1m \\
\cmidrule{1-8} 
\textit{\textbf{MiAP scores:}} \\
\rr UserTags              & 0.204  & 0.204 & 0.204 && 0.255 & 0.255 & 0.255 \\ [2pt]
\rr TagCooccur            & 0.213  & 0.242 & 0.253 && 0.269 & 0.305 & 0.317 \\ [2pt]
\rr BovW + TagCooccur+    & 0.217 & 0.262 & 0.286  && 0.245 & 0.297 & 0.324 \\
\rr BovW + RobustPCA      & 0.271 & \textbf{0.310} & -- & & \textbf{0.332} & 0.323 & -- \\
\rr BovW + TensorAnalysis & \textsuperscript{*}0.298 & \textsuperscript{*}0.297 & -- & & -- & -- & -- \\ [3pt]

\rr CNN + TagCooccur+     & 0.234 & 0.277 & 0.310           && 0.305 & 0.359 & 0.387 \\
\rr CNN + RobustPCA       & 0.368 & \textbf{0.376} & --     && \textbf{0.424} & 0.419 & -- \\ 
\rr CNN + TensorAnalysis  & --    & --     & --             && -- & -- & -- \\ [3pt]
\textit{\textbf{MAP scores:}} \\
\rr UserTags              & 0.263 & 0.263 & 0.263          && 0.338 & 0.338 & 0.338 \\ [2pt]
\rr TagCooccur            & 0.266 & 0.298 & 0.313          && 0.223 & 0.321 & 0.308 \\ [2pt]
\rr BovW + TagCooccur+    & 0.294 & 0.343 & \textbf{0.377} && 0.231 & 0.345 & \textbf{0.353} \\
\rr BovW + RobustPCA      & 0.225 & 0.337 & --             && 0.229 & 0.234 & -- \\ 
\rr BovW + TensorAnalysis & --    & --     & --             && -- & -- & -- \\ [3pt]

\rr CNN + TagCooccur+     & 0.330 & 0.381 & 0.420          && 0.264 & 0.391 & 0.406 \\ 
\rr CNN + RobustPCA       & 0.566 & \textbf{0.627} & --    && 0.439 & \textbf{0.440} & -- \\ 
\rr CNN + TensorAnalysis  & --    & --     & --             && -- & -- & -- \\ [3pt]
\bottomrule
\end{tabular}
}
}
\end{table}

\begin{figure}[!tb]
\centering
                \includegraphics[width=1.\textwidth]{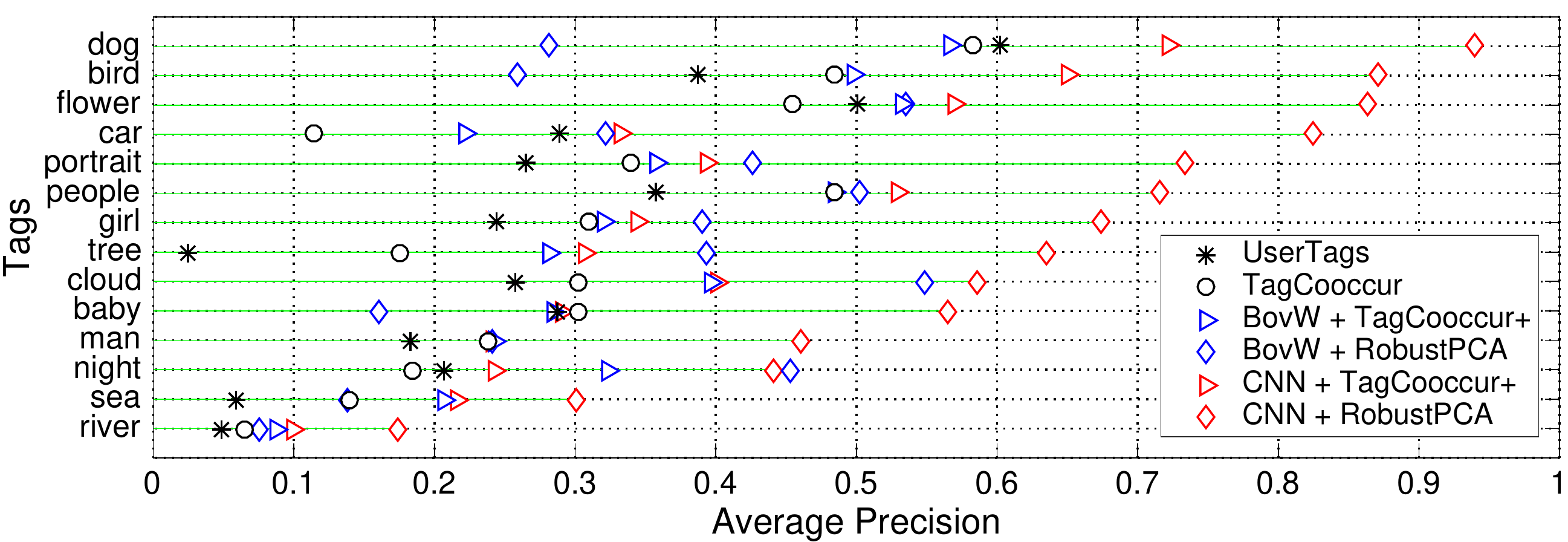}
\caption{
\textbf{Per-tag comparison of methods for tag refinement on MIRFlickr}, trained on Train100k. The colors identify the features used: \textbf{blue} for BovW, \textbf{red} for CNN. The test tags have been sorted in descending order by the performance of CNN + RobustPCA.
} 
\label{fig:tag_comparison_mirflickr_refinement}
\end{figure}

RobustPCA outperforms the competitors on both test sets, when provided with the CNN feature. 
Fig. \ref{fig:tag_comparison_mirflickr_refinement} presents a per-tag comparison on MIRFlickr.
RobustPCA has the best scores for 9 out of the 14 tags with BovW, and wins all the tags when CNN is used. 

Concerning the influence of the media dimension, the tag + image based methods (RobustPCA and TagCooccur+) are
in general better than the tag based method (TagCooccur).
As shown in Fig. \ref{fig:tag_comparison_mirflickr_refinement}, 
except for 3 out of 14 MIRFlickr test tags with BovW, using the image media is beneficial.
As in the tag assignment task, the use of the CNN feature strongly improves the performance.

Concerning the learning methods, TensorAnalysis has the potential to leverage tag, image, and user simultaneously. However, due to its relatively poor scalability, we were able to run this method only with Train10k and Train100k on MIRFlickr. For Train10k, TensorAnalysis yielded higher MiAP than RobustPCA, probably thanks to its capability of modeling user correlations. It is outperformed by RobustPCA when more training data is used.

As more training data is used, the performance of TagCooccur, TagCooccur+, and RobustPCA on MIRFlickr consistently improves.
Since these three methods rely on data-driven tag affinity, image affinity, or tag and image affinity, a small set of 10k images is generally inadequate to compute these affinities. 
The effect of increasing the training set size is clearly visible if we compare scores corresponding to Train10k and Train100k. 
The results on NUS-WIDE show some inconsistency.
For TagCooccur, MiAP improves from Train100k to Train1m,  while MAP drops.  
This is presumably due to the fact that in the experiments we used the parameters recommended in the original paper,  appropriately selected to optimize tag ranking. Hence, they might be suboptimal for image ranking.
BovW + RobustPCA scores a lower MAP than BovW + TagCooccur+. 
This is probably due to the fact that the low-rank matrix factorization technique, while being able to jointly exploit tag and image information, is more sensitive to the content-based representation.

\begin{figure}[!tb]
\centering
                \includegraphics[width=1.\textwidth]{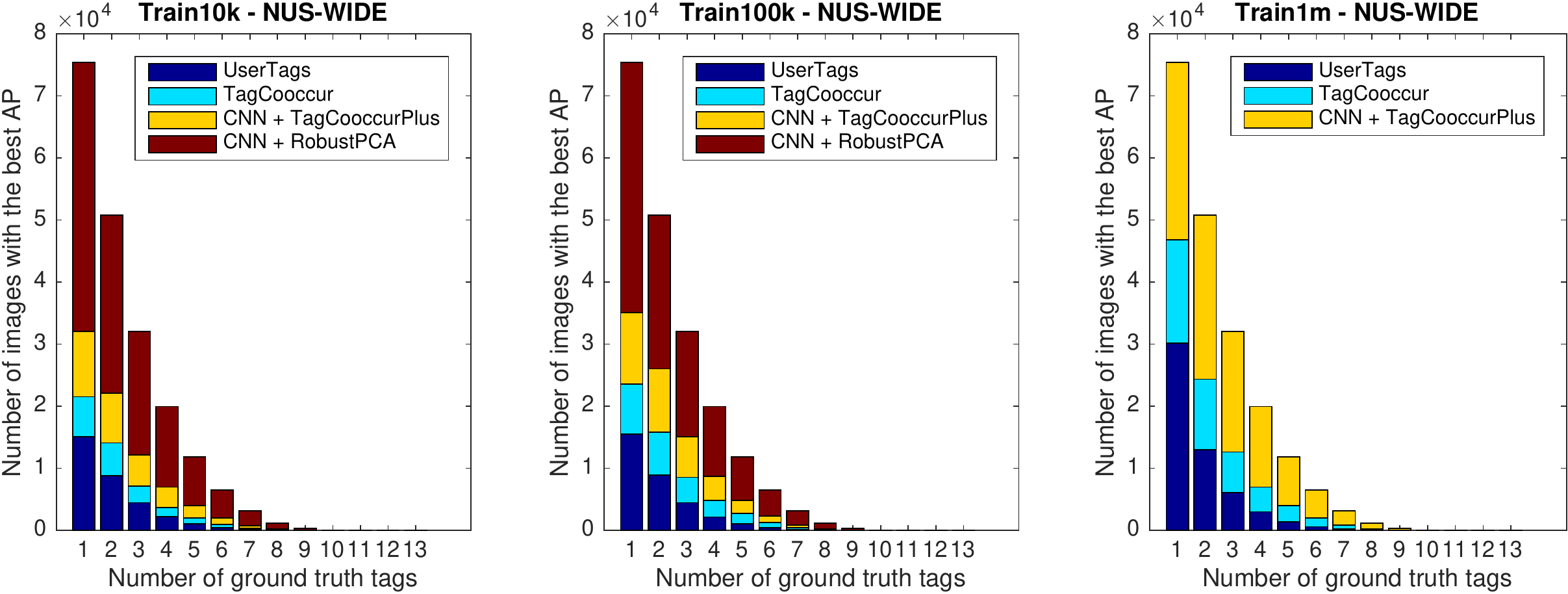}
\caption{
\textbf{Per-image comparison of methods for tag refinement on NUS-WIDE}. Test images are grouped in terms of their number of ground truth tags. The area of a colored bar is proportional to the number of images that the corresponding method scores best.}
\label{fig:miap_comparison_nuswide_refinement}
\end{figure}

A per-image comparison is given in Fig. \ref{fig:miap_comparison_nuswide_refinement}.
As for tag assignment, the test images have been grouped according to the number of ground truth tags associated. The size of the colored areas is proportional to the number of images where the corresponding method scores best. For the majority of test image, the three tag refinement methods have higher average precision than UserTags.
This means more relevant tags are added, so the tags are refined.
It should be noted that the success of tag refinement depends much on the quality of the original tags assigned to the test images. 
Examples are shown in Table \ref{tab:tagging-result}: in row 6, although the tag `earthquake' is irrelevant to the image content, it is ranked at the top by RobustPCA.
To what extent a tag refinement method shall count on the existing tags is tricky.

To summarize, the tag + image based methods outperform the tag based method for
tag refinement. 
RobustPCA is the best, and improves as more training data is employed. 
Nonetheless,
implementing RobustPCA is challenging for both computation and memory footprint. In contrast, TagCooccur+ is more scalable and it can learn from large-scale data.

\begin{table} [tbh!]
\renewcommand{\arraystretch}{1}
\tbl{Selected tag assignment and refinement results on NUS-WIDE. Visual feature: BovW. The top five ranked tags are shown, with correct prediction marked by the \textit{\textbf{bold italic}} font. \label{tab:tagging-result}
}
{\centering
\scalebox{0.81}{
\begin{tabular}{@{}lll llll lll@{}}
\toprule
           &              &           & \multicolumn{4}{c}{\textbf{Tag assignment}} & \multicolumn{3}{c}{\textbf{Tag refinement}} \\ 
           \cmidrule(lr){4-7} \cmidrule(lr){8-10} 
Test image & Ground truth & User tags & KNN & TagVote & TagProp & RelExample & TagCooccur & TagCooccur+ & RobustPCA \\

\cmidrule{1-10}
\tabincell{l}{\includegraphics[width=2.15cm,height=2.15cm]{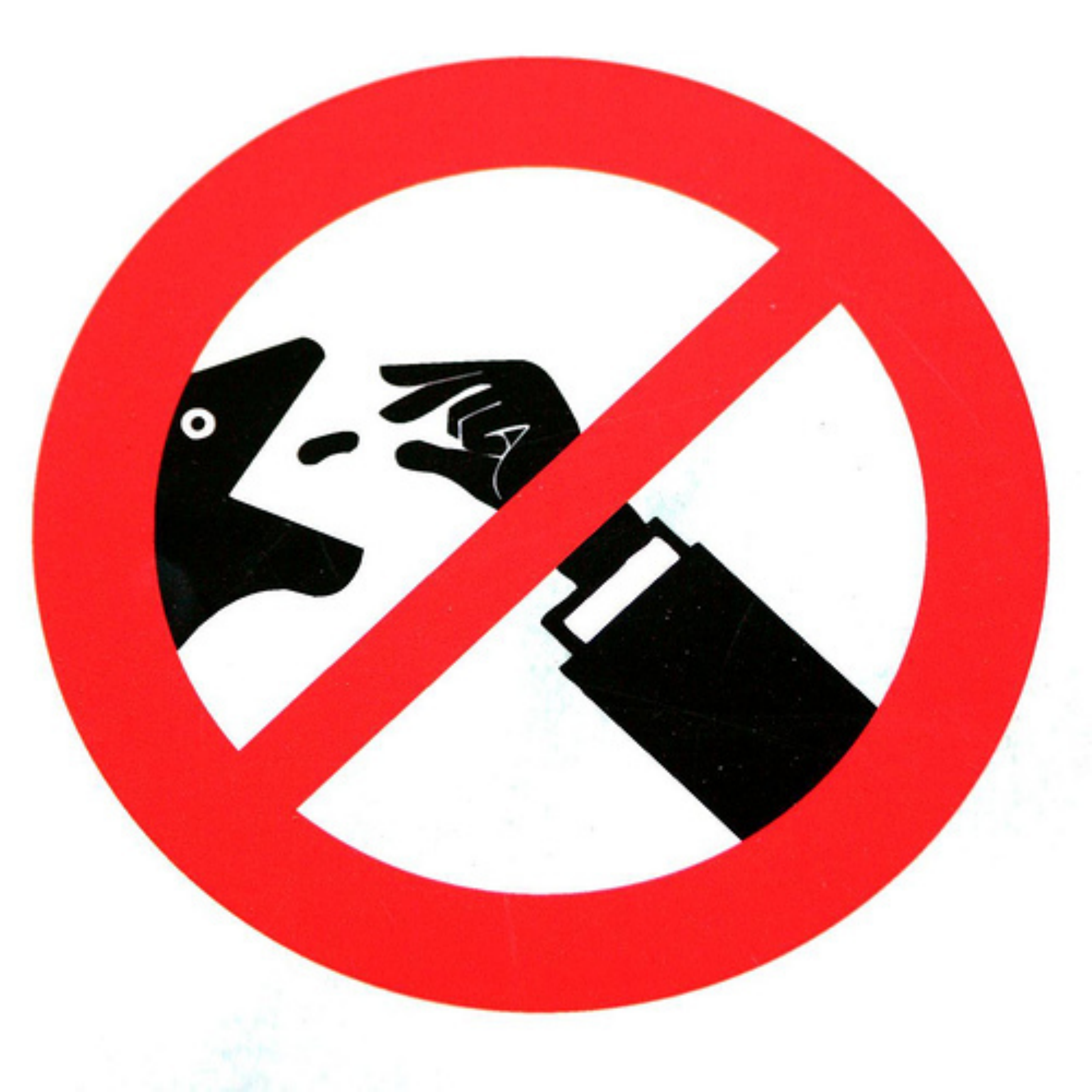}}&\tabincell{l}{sign}&\tabincell{l}{\black{\textit{\textbf{sign}}}\\reptile\\zoo\\red\\white}&\tabincell{l}{animal\\flower\\car\\horse\\street}&\tabincell{l}{dog\\house\\bird\\\black{\textit{\textbf{sign}}}\\bear}&\tabincell{l}{\black{\textit{\textbf{sign}}}\\street\\flower\\dog\\bird}&\tabincell{l}{soccer\\whale\\book\\toy\\moon}&\tabincell{l}{animal\\street\\\black{\textit{\textbf{sign}}}\\water\\car}&\tabincell{l}{\black{\textit{\textbf{sign}}}\\bird\\dog\\animal\\toy}&\tabincell{l}{\black{\textit{\textbf{sign}}}\\bird\\flower\\animal\\street}\\
\cmidrule{1-10}
\tabincell{l}{\includegraphics[width=2.15cm,height=2.15cm]{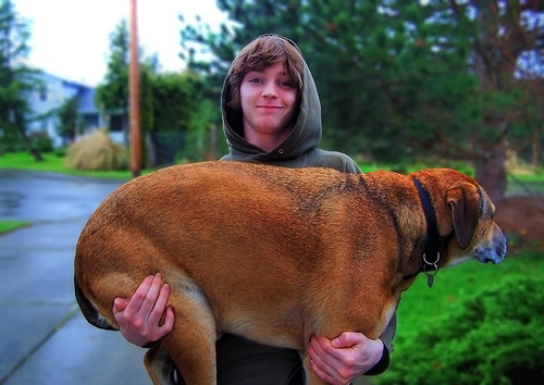}}&\tabincell{l}{animal\\dog\\person}&\tabincell{l}{colour\\color\\\black{\textit{\textbf{dog}}}\\hound}&\tabincell{l}{flower\\garden\\horse\\tree\\\black{\textit{\textbf{dog}}}}&\tabincell{l}{garden\\flower\\food\\cat\\\black{\textit{\textbf{dog}}}}&\tabincell{l}{flower\\\black{\textit{\textbf{dog}}}\\garden\\car\\tree}&\tabincell{l}{garden\\\black{\textit{\textbf{dog}}}\\fish\\fox\\\black{\textit{\textbf{animal}}}}&\tabincell{l}{\black{\textit{\textbf{dog}}}\\\black{\textit{\textbf{animal}}}\\car\\beach\\flower}&\tabincell{l}{\black{\textit{\textbf{dog}}}\\flower\\\black{\textit{\textbf{animal}}}\\cat\\food}&\tabincell{l}{\black{\textit{\textbf{dog}}}\\flower\\\black{\textit{\textbf{animal}}}\\water\\garden}\\
\cmidrule{1-10}
\tabincell{l}{\includegraphics[width=2.15cm,height=2.15cm]{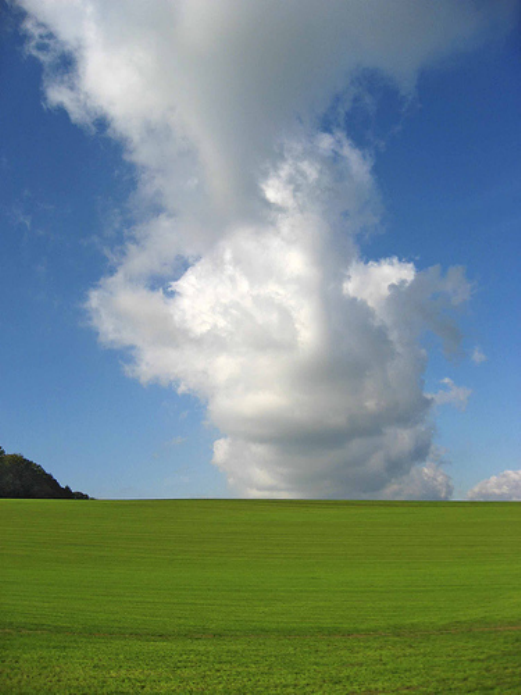}}&\tabincell{l}{cloud\\grass\\sky}&\tabincell{l}{\black{\textit{\textbf{cloud}}}\\\black{\textit{\textbf{grass}}}}&\tabincell{l}{\black{\textit{\textbf{cloud}}}\\\black{\textit{\textbf{sky}}}\\beach\\water\\snow}&\tabincell{l}{\black{\textit{\textbf{cloud}}}\\\black{\textit{\textbf{sky}}}\\water\\beach\\mountain}&\tabincell{l}{\black{\textit{\textbf{cloud}}}\\\black{\textit{\textbf{sky}}}\\beach\\water\\lake}&\tabincell{l}{\black{\textit{\textbf{cloud}}}\\ocean\\surf\\\black{\textit{\textbf{sky}}}\\beach}&\tabincell{l}{\black{\textit{\textbf{grass}}}\\\black{\textit{\textbf{sky}}}\\tree\\flower\\water}&\tabincell{l}{\black{\textit{\textbf{cloud}}}\\\black{\textit{\textbf{sky}}}\\water\\beach\\tree}&\tabincell{l}{\black{\textit{\textbf{cloud}}}\\\black{\textit{\textbf{grass}}}\\\black{\textit{\textbf{sky}}}\\water\\mountain}\\
\cmidrule{1-10}
\tabincell{l}{\includegraphics[width=2.15cm,height=2.15cm]{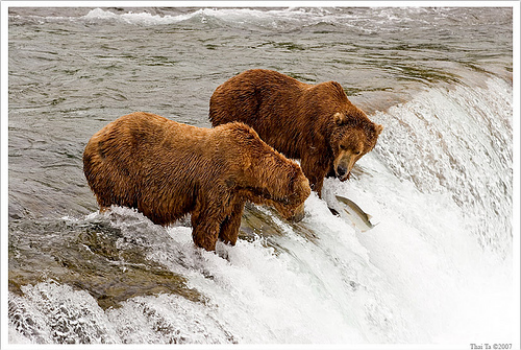}}&\tabincell{l}{animal\\bear\\water}&\tabincell{l}{brown\\\black{\textit{\textbf{bear}}}\\salmon\\national\\park}&\tabincell{l}{snow\\beach\\\black{\textit{\textbf{animal}}}\\\black{\textit{\textbf{water}}}\\tree}&\tabincell{l}{snow\\\black{\textit{\textbf{animal}}}\\waterfall\\tree\\\black{\textit{\textbf{water}}}}&\tabincell{l}{snow\\beach\\sand\\\black{\textit{\textbf{bear}}}\\\black{\textit{\textbf{water}}}}&\tabincell{l}{\black{\textit{\textbf{water}}}\\sand\\rock\\surf\\ocean}&\tabincell{l}{waterfall\\\black{\textit{\textbf{water}}}\\tree\\\black{\textit{\textbf{bear}}}\\\black{\textit{\textbf{animal}}}}&\tabincell{l}{waterfall\\\black{\textit{\textbf{water}}}\\\black{\textit{\textbf{animal}}}\\snow\\tree}&\tabincell{l}{\black{\textit{\textbf{water}}}\\waterfall\\\black{\textit{\textbf{bear}}}\\\black{\textit{\textbf{animal}}}\\snow}\\
\cmidrule{1-10}
\tabincell{l}{\includegraphics[width=2.15cm,height=2.15cm]{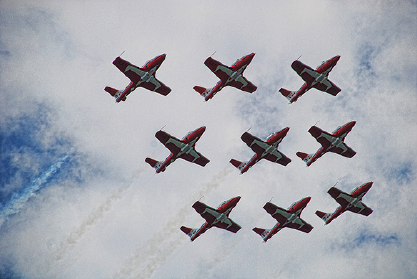}}&\tabincell{l}{airplane\\cloud\\military\\sky}&\tabincell{l}{flag\\great}&\tabincell{l}{\black{\textit{\textbf{sky}}}\\\black{\textit{\textbf{cloud}}}\\snow\\bird\\\black{\textit{\textbf{airplane}}}}&\tabincell{l}{snow\\\black{\textit{\textbf{cloud}}}\\\black{\textit{\textbf{sky}}}\\mountain\\bird}&\tabincell{l}{\black{\textit{\textbf{airplane}}}\\\black{\textit{\textbf{sky}}}\\snow\\bird\\airport}&\tabincell{l}{snow\\frost\\bird\\\black{\textit{\textbf{airplane}}}\\tattoo}&\tabincell{l}{car\\street\\snow\\water\\\black{\textit{\textbf{sky}}}}&\tabincell{l}{snow\\\black{\textit{\textbf{sky}}}\\\black{\textit{\textbf{cloud}}}\\mountain\\bird}&\tabincell{l}{flag\\\black{\textit{\textbf{sky}}}\\snow\\\black{\textit{\textbf{cloud}}}\\bird}\\
\cmidrule{1-10}
\tabincell{l}{\includegraphics[width=2.15cm,height=2.15cm]{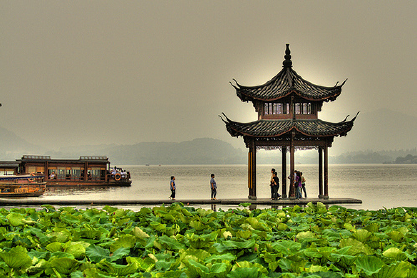}}&\tabincell{l}{cloud\\garden\\sky\\water}&\tabincell{l}{china\\earthquake\\people\\hangzhou\\summer\\westlake}&\tabincell{l}{car\\beach\\\black{\textit{\textbf{water}}}\\street\\tree}&\tabincell{l}{grass\\tree\\\black{\textit{\textbf{water}}}\\road\\bridge}&\tabincell{l}{car\\road\\street\\\black{\textit{\textbf{sky}}}\\bird}&\tabincell{l}{house\\road\\grass\\bird\\sand}&\tabincell{l}{\black{\textit{\textbf{water}}}\\flower\\street\\temple\\tree}&\tabincell{l}{tree\\\black{\textit{\textbf{water}}}\\street\\\black{\textit{\textbf{garden}}}\\car}&\tabincell{l}{earthquake\\\black{\textit{\textbf{water}}}\\tree\\\black{\textit{\textbf{cloud}}}\\\black{\textit{\textbf{sky}}}}\\
\cmidrule{1-10}
\tabincell{l}{\includegraphics[width=2.15cm,height=2.15cm]{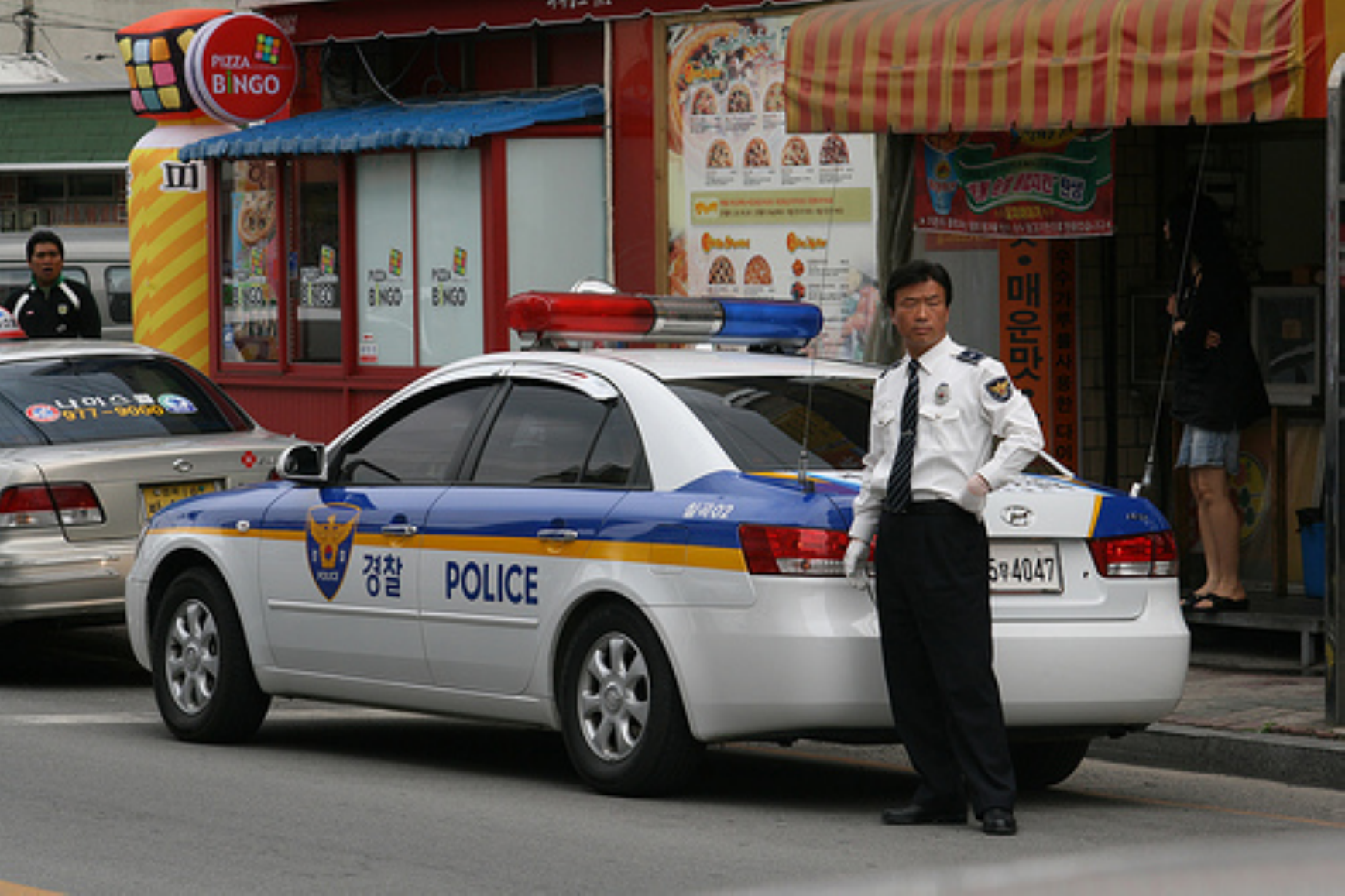}}&\tabincell{l}{police\\road\\vehicle\\window}&\tabincell{l}{farmer\\dog\\motorcycle\\\black{\textit{\textbf{police}}}\\train}&\tabincell{l}{car\\street\\\black{\textit{\textbf{police}}}\\\black{\textit{\textbf{vehicle}}}\\\black{\textit{\textbf{road}}}}&\tabincell{l}{car\\street\\\black{\textit{\textbf{police}}}\\\black{\textit{\textbf{vehicle}}}\\sport}&\tabincell{l}{\black{\textit{\textbf{police}}}\\car\\street\\\black{\textit{\textbf{road}}}\\sport}&\tabincell{l}{\black{\textit{\textbf{police}}}\\\black{\textit{\textbf{vehicle}}}\\street\\car\\sport}&\tabincell{l}{street\\car\\animal\\train\\bird}&\tabincell{l}{car\\street\\\black{\textit{\textbf{police}}}\\food\\horse}&\tabincell{l}{\black{\textit{\textbf{police}}}\\train\\dog\\bird\\car}\\
\cmidrule{1-10}

\tabincell{l}{\includegraphics[width=2.15cm,height=2.15cm]{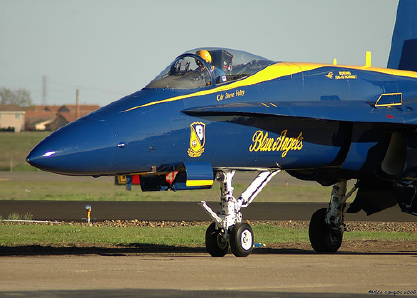}}&\tabincell{l}{airplane\\airport\\cloud\\military\\sky\\vehicle}&\tabincell{l}{vertical\\sunglass\\smoke\\pilot\\landing}&\tabincell{l}{car\\beach\\street\\water\\\black{\textit{\textbf{airplane}}}}&\tabincell{l}{car\\street\\sport\\\black{\textit{\textbf{airplane}}}\\\black{\textit{\textbf{vehicle}}}}&\tabincell{l}{car\\sport\\\black{\textit{\textbf{airplane}}}\\\black{\textit{\textbf{vehicle}}}\\road}&\tabincell{l}{\black{\textit{\textbf{airplane}}}\\sport\\\black{\textit{\textbf{airport}}}\\\black{\textit{\textbf{vehicle}}}\\\black{\textit{\textbf{military}}}}&\tabincell{l}{\black{\textit{\textbf{airplane}}}\\car\\\black{\textit{\textbf{military}}}\\\black{\textit{\textbf{airport}}}\\street}&\tabincell{l}{car\\\black{\textit{\textbf{airplane}}}\\street\\\black{\textit{\textbf{airport}}}\\\black{\textit{\textbf{military}}}}&\tabincell{l}{\black{\textit{\textbf{airplane}}}\\car\\\black{\textit{\textbf{sky}}}\\\black{\textit{\textbf{cloud}}}\\water}\\
\cmidrule{1-10}
\tabincell{l}{\includegraphics[width=2.15cm,height=2.15cm]{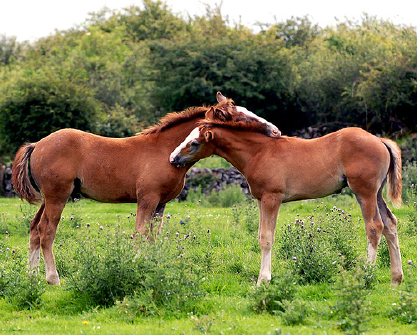}}&\tabincell{l}{animal\\grass\\horse}&\tabincell{l}{\black{\textit{\textbf{horse}}}\\pony\\run\\field\\brown}&\tabincell{l}{waterfall\\tree\\garden\\water\\\black{\textit{\textbf{horse}}}}&\tabincell{l}{\black{\textit{\textbf{animal}}}\\tree\\\black{\textit{\textbf{horse}}}\\garden\\waterfall}&\tabincell{l}{garden\\\black{\textit{\textbf{grass}}}\\\black{\textit{\textbf{horse}}}\\tree\\waterfall}&\tabincell{l}{cow\\elk\\\black{\textit{\textbf{animal}}}\\\black{\textit{\textbf{grass}}}\\\black{\textit{\textbf{horse}}}}&\tabincell{l}{\black{\textit{\textbf{horse}}}\\car\\\black{\textit{\textbf{animal}}}\\street\\dog}&\tabincell{l}{\black{\textit{\textbf{animal}}}\\\black{\textit{\textbf{horse}}}\\tree\\dog\\water}&\tabincell{l}{\black{\textit{\textbf{animal}}}\\\black{\textit{\textbf{horse}}}\\tree\\water\\flower}\\

\bottomrule
\end{tabular}
}}
\end{table}

\subsection{Tag retrieval} \label{ssec:exp-retrieval}

Table \ref{tab:exp-retrieval} shows the performance of different methods for tag retrieval.
Recall that when retrieving images for a specific test tag, we consider only images that are labeled with this tag.
Hence, MAP scores here are higher than their counterpart in Table \ref{tab:exp-tagrefine}.

We start our analysis by comparing the three baselines, namely UserTags, TagNum, and TagPosition, which retrieve images simply by the original tags.
As it can be noticed, TagNum and TagPosition are more effective than UserTags, TagNum outperforms TagPosition on Flickr51, and the latter has better scores on NUS-WIDE.
The effectiveness of such metadata based features depend much on datasets, and are unreliable for tag retrieval.

All the methods considered have higher MAP than the three baselines.
All the methods have better performance than the baselines on Flickr51 and performance increases with the size of the training set. 
On NUS-WIDE, SemanticField, TagCooccur, and TagRanking, are less effective than TagPosition. 
We attribute this result to the fact that, for these methods, the tag relevance functions favor images with fewer tags. 
So they closely follow similar performance and dataset dependency. 

\begin{table} [tb!]
\renewcommand{\arraystretch}{1.2}
\tbl{Evaluating methods for tag retrieval. 
Given the same feature, bold values indicate top performers on individual test sets per performance metric.
\label{tab:exp-retrieval}}
{\centering
\scalebox{0.93}{
\begin{tabular}{@{}l r r r r r r r@{}}
\toprule
                & \multicolumn{3}{c}{\textbf{Flickr51}} && \multicolumn{3}{c}{\textbf{NUS-WIDE}}  \\
                \cmidrule(lr){2-4} \cmidrule(l){6-8} 
\textbf{Method} & Train10k &  Train100k & Train1m && Train10k &  Train100k & Train1m \\
\cmidrule{1-8} 
\textit{\textbf{MAP scores:}} \\
\rr UserTags     & 0.595     & 0.595  & 0.595     && 0.489 & 0.489     & 0.489 \\
\rr TagNum       & 0.664     & 0.664  & 0.664     && 0.520 & 0.520     & 0.520 \\ 
\rr TagPosition  & 0.640     & 0.640  & 0.640     && 0.557 & 0.557     & 0.557 \\ [3pt]

\rr SemanticField      & 0.687     & 0.707  & 0.713     && 0.565 & 0.584 & 0.584 \\
\rr TagCooccur         & 0.625     & 0.679  & 0.704     && 0.534 & 0.576 & 0.588 \\ [3pt]

\rr BovW + TagCooccur+ & 0.640     & 0.732  & 0.764     && 0.560 & 0.622 & 0.643 \\
\rr BovW + TagRanking  & 0.685     & 0.686  & 0.708     && 0.557 & 0.574 & 0.578 \\
\rr BovW + KNN         & 0.678     & 0.742  & 0.770     && 0.587 & 0.632 & 0.658 \\
\rr BovW + TagVote     & 0.678     & 0.741  & 0.769     && 0.587 & 0.632 & 0.659 \\
\rr BovW + TagProp     & 0.671     & 0.748  & 0.772     && 0.585 & 0.636 & 0.657 \\
\rr BovW + TagFeature  & 0.689     & 0.726  & 0.737     && 0.589 & 0.602 & 0.606 \\
\rr BovW + RelExample  & 0.706     & 0.756  & \textbf{0.783}    && 0.609 & 0.645     & \textbf{0.663}  \\ 
\rr BovW + RobustPCA   & 0.697     & 0.701  & --        && 0.650 & 0.650 & -- \\ 
\rr BovW + TensorAnalysis & --     & --     & --        && --    & --        & --  \\ [3pt]

\rr CNN + TagCooccur+  & 0.654     & 0.781  & 0.821     && 0.572 & 0.653 & 0.674 \\
\rr CNN + TagRanking   & 0.744     & 0.735  & 0.747     && 0.589 & 0.590 & 0.590 \\
\rr CNN + KNN          & 0.811     & 0.859  & 0.880     && 0.683 & 0.722 & 0.734 \\
\rr CNN + TagVote      & 0.808     & 0.859  & \textbf{0.881}  && 0.675 & 0.724 & \textbf{0.738} \\
\rr CNN + TagProp      & 0.824     & 0.867  & 0.879     && 0.689 & 0.727 & 0.731 \\
\rr CNN + TagFeature   & 0.827     & 0.853  & 0.859     && 0.675 & 0.700 & 0.703 \\
\rr CNN + RelExample   & 0.838     & 0.863  & 0.878     && 0.689 & 0.717 & 0.734\\
\rr CNN + RobustPCA    & 0.811     & 0.839  & --        && 0.725 & 0.726 & -- \\ 
\rr CNN + TensorAnalysis & --     & --     & --        && --    & --        & --  \\ [3pt]

\textit{\textbf{NDCG$_{20}$ scores:}} \\
\rr UserTags     & 0.432     & 0.432  & 0.432     && 0.487 & 0.487     & 0.487 \\
\rr TagNum       & 0.522     & 0.522  & 0.522     && 0.541 & 0.541     & 0.541 \\ 
\rr TagPosition  & 0.511     & 0.511  & 0.511     && 0.623 & 0.623     & 0.623 \\ [3pt]

\rr SemanticField      & 0.591     & 0.623  & 0.645     && 0.596 & 0.622 & 0.624 \\    
\rr TagCooccur         & 0.482     & 0.527  & 0.631     && 0.529 & 0.602 & 0.614 \\ [3pt]

\rr BovW + TagCooccur+ & 0.503     & 0.625  & 0.686     && 0.590 & 0.681 & 0.734 \\
\rr BovW + TagRanking  & 0.530     & 0.568  & 0.571     && 0.557 & 0.572     & 0.572  \\
\rr BovW + KNN         & 0.577     & 0.699  & 0.756     && 0.638 & 0.734     & 0.799 \\
\rr BovW + TagVote     & 0.573     & 0.701  & 0.754     && 0.629 & 0.734     & 0.804 \\
\rr BovW + TagProp     & 0.570     & 0.715  & \textbf{0.759}     && 0.666 & 0.750     & \textbf{0.809} \\
\rr BovW + TagFeature  & 0.547     & 0.626  & 0.646     && 0.622 & 0.615     & 0.618  \\
\rr BovW + RelExample  & 0.614     & 0.722  & 0.748     && 0.692 & 0.736     & 0.776  \\
\rr BovW + RobustPCA   & 0.549     & 0.548  & --        && 0.768 & 0.781     & -- \\ 
\rr BovW + TensorAnalysis & --     & --     & --        && --    & --        & --  \\ [3pt]

\rr CNN + TagCooccur+  & 0.504     & 0.615  & 0.724     && 0.571 & 0.705 & 0.738 \\
\rr CNN + TagRanking   & 0.577     & 0.607  & 0.597     && 0.578 & 0.594 & 0.583 \\
\rr CNN + KNN          & 0.709     & 0.830  & 0.897     && 0.773 & 0.832 & 0.863 \\
\rr CNN + TagVote      & 0.722     & 0.826  & \textbf{0.899}   && 0.740  & 0.837 & \textbf{0.879} \\
\rr CNN + TagProp      & 0.768     & 0.857  & 0.865     && 0.764 & 0.839 & 0.845 \\
\rr CNN + TagFeature   & 0.755     & 0.813  & 0.818     && 0.704 & 0.807 & 0.787 \\
\rr CNN + RelExample   & 0.764     & 0.843  & 0.879     && 0.773 & 0.814 & 0.866 \\
\rr CNN + RobustPCA    & 0.733     & 0.821  & --        && 0.865 & 0.862 & -- \\ 
\rr CNN + TensorAnalysis & --     & --     & --        && --    & --        & --  \\

\bottomrule
\end{tabular}
}}
\end{table}

\begin{figure}[!bt]
\centering
                \includegraphics[width=\textwidth]{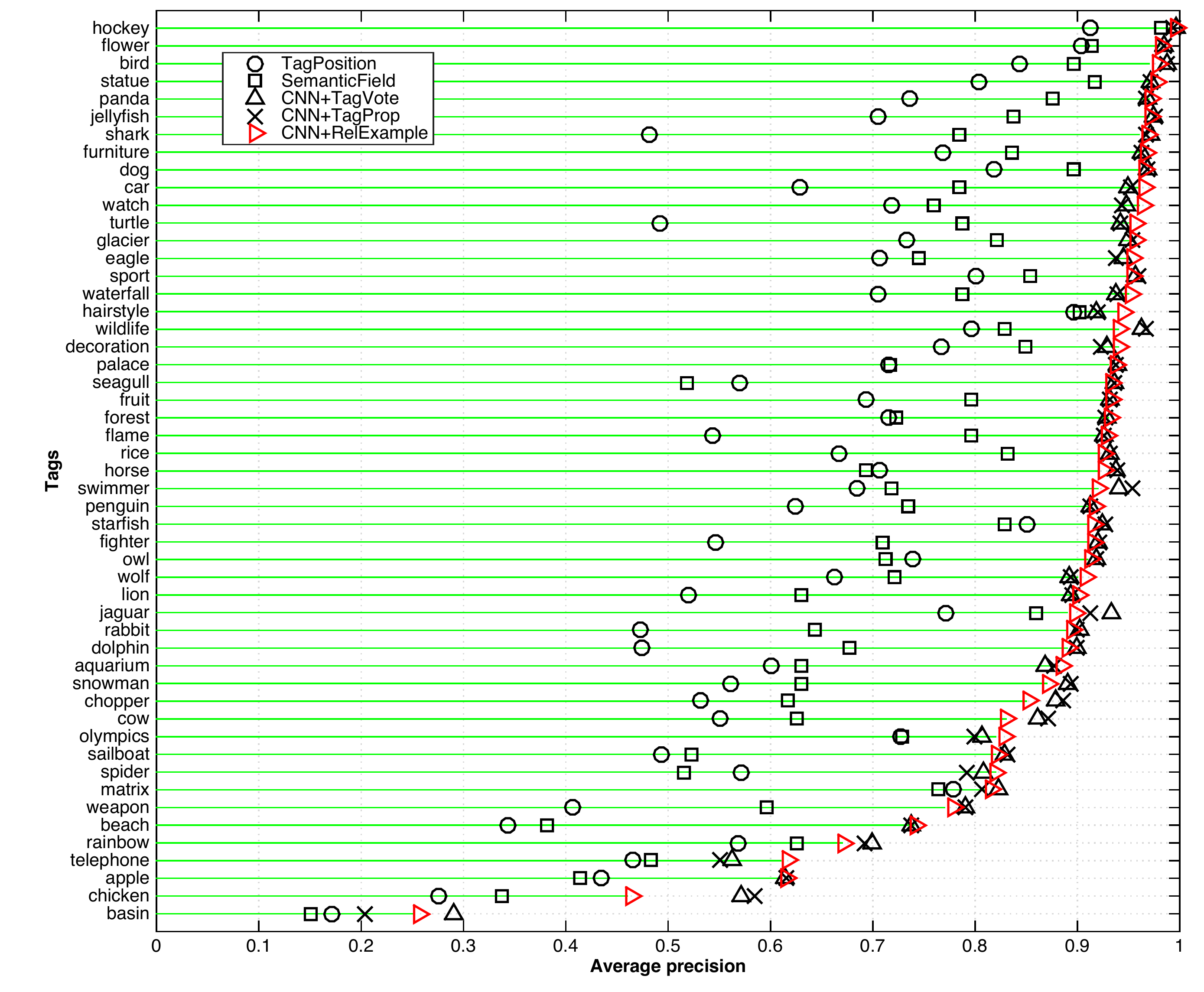}
\caption{
\textbf{Per-tag comparison between TagPosition, SemanticField, TagVote, TagProp, and RelExample on Flickr51},
with Train1m as the training set. The 51 test tags have been sorted in descending order by the performance of RelExample. 
} \label{fig:flickr55ap}
\end{figure}

\begin{figure}[!bt]
\centering
\subfigure[TagPosition]{
                \includegraphics[width=0.22\textwidth]{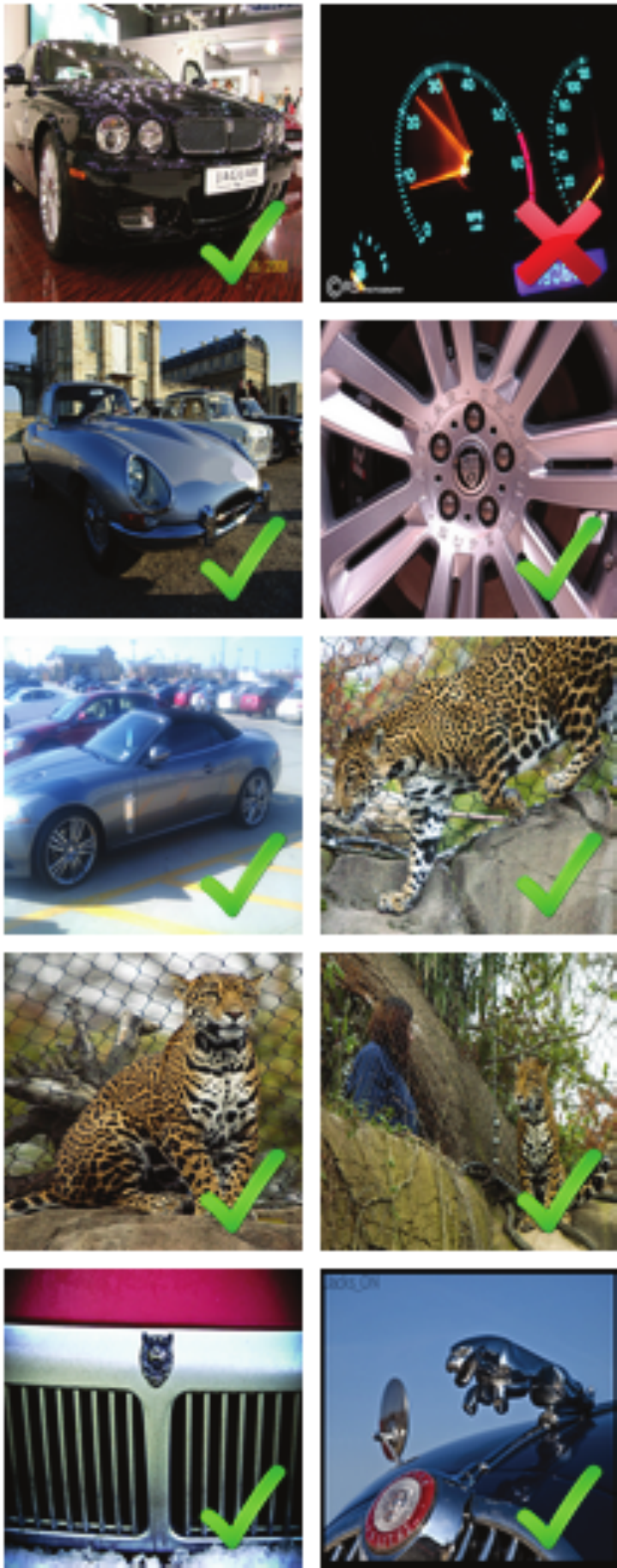}
                \label{fig:jaguar-tp}
}
\subfigure[SemanticField]{
                \includegraphics[width=0.22\textwidth]{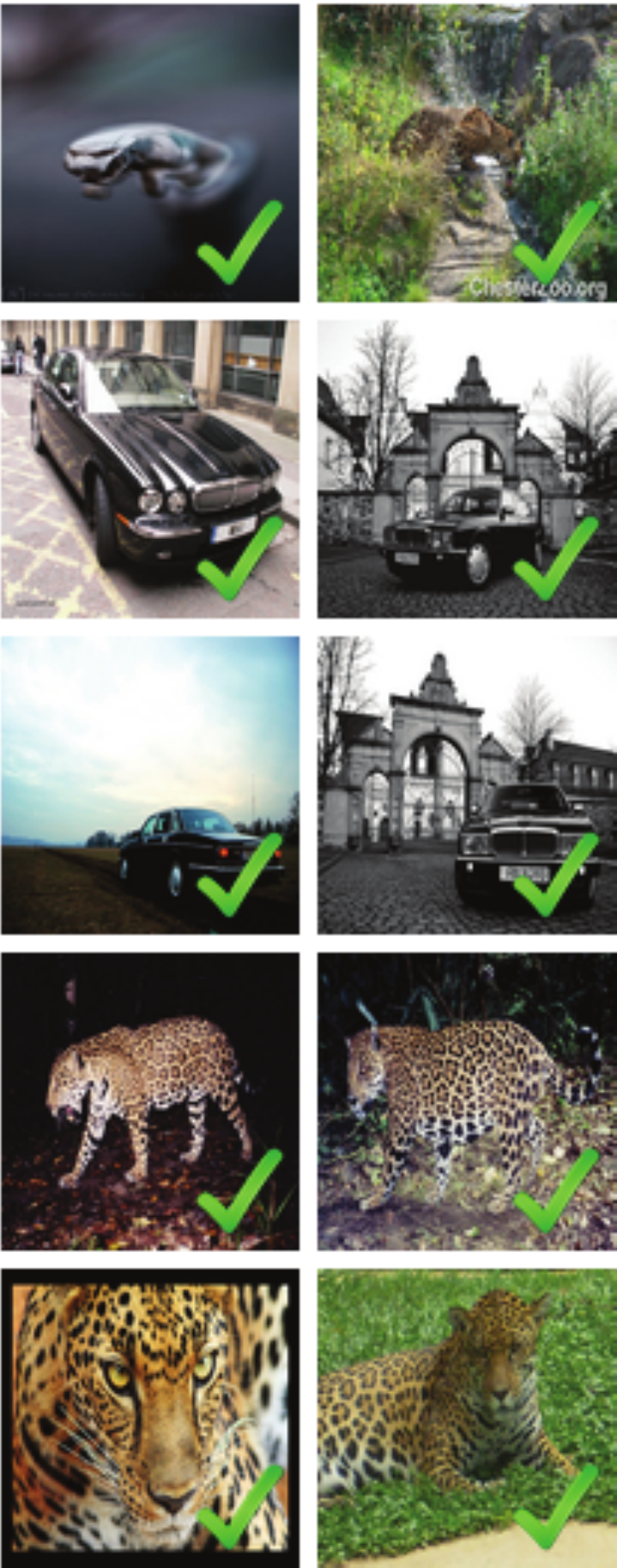}
                \label{fig:jaguar-sf}
}
\subfigure[BovW + RelExample]{
                \includegraphics[width=0.22\textwidth]{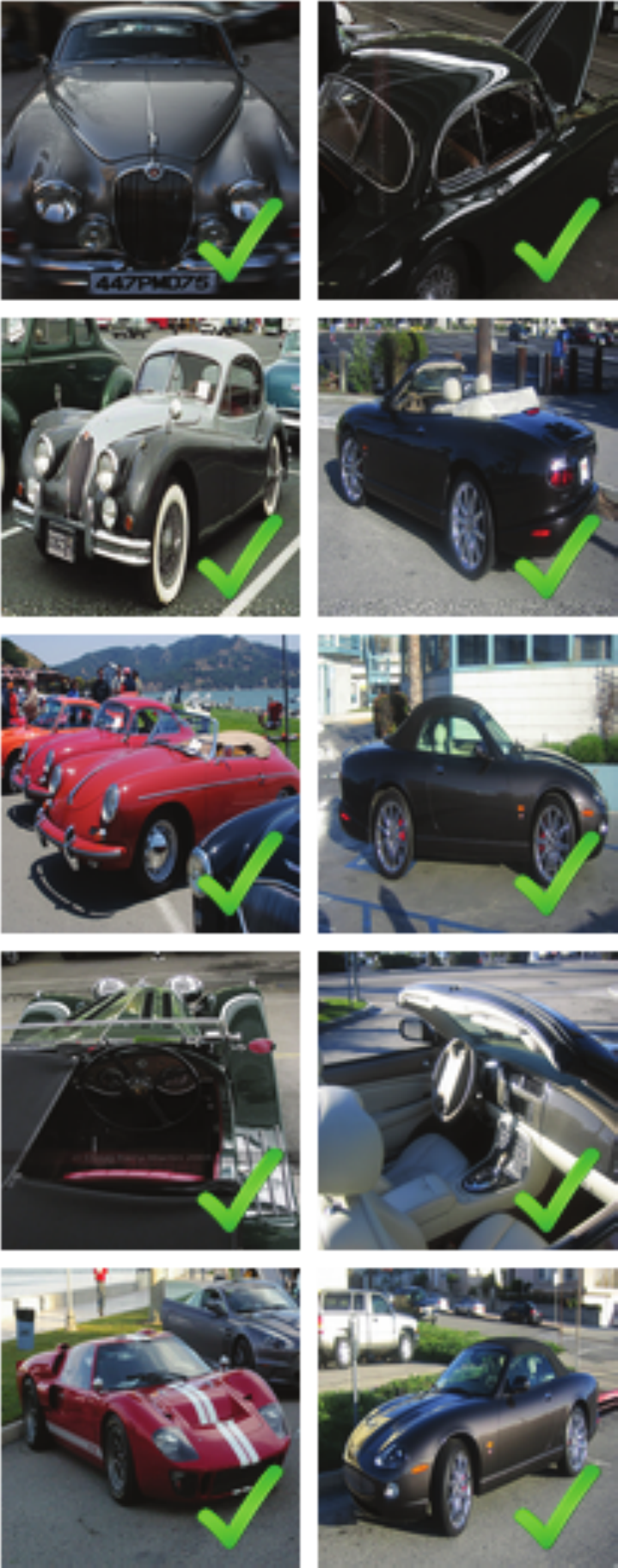}
                \label{fig:jaguar-bovw-fiksvm}
}
\subfigure[CNN + RelExample]{
                \includegraphics[width=0.22\textwidth]{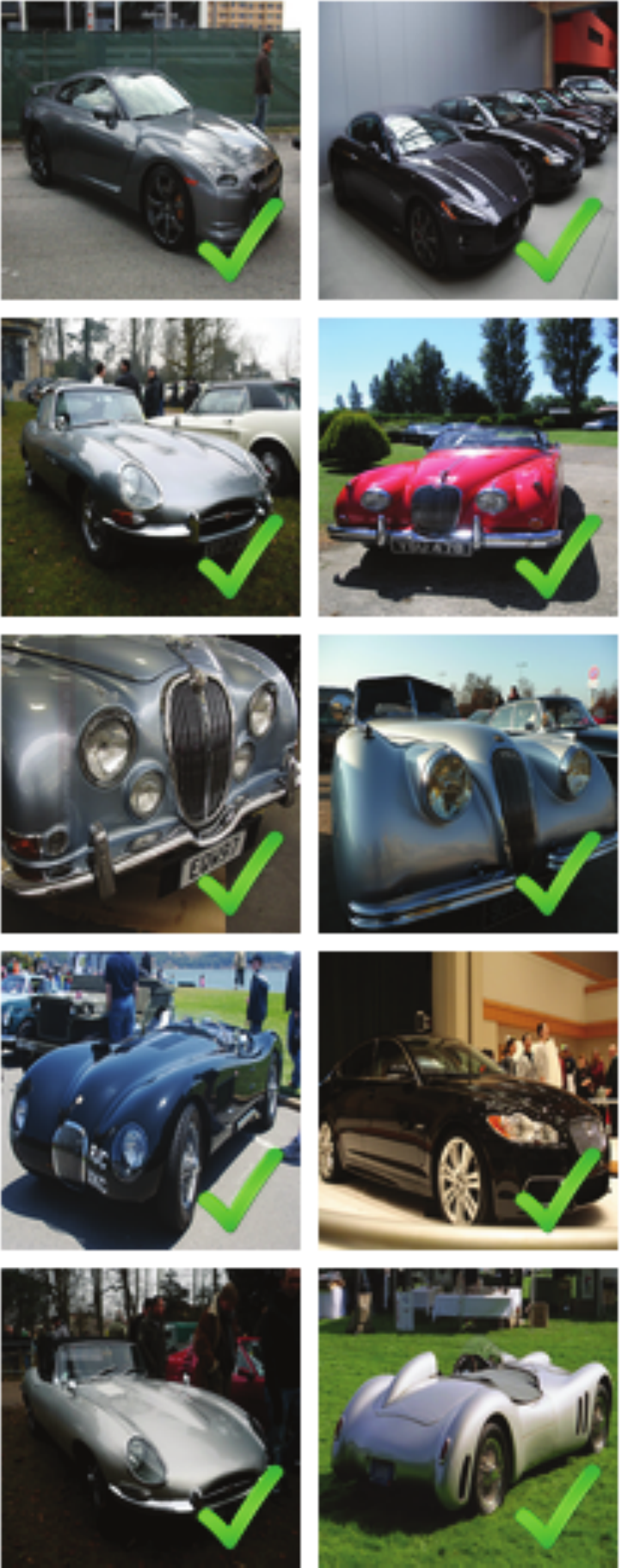}
                \label{fig:jaguar-cnn-fiksvm}
}
\caption{
\textbf{Top 10 ranked images of `jaguar', by (a) TagPosition, (b) SemanticField, (c) BovW + RelExample, and (d) CNN + RelExample}. Checkmarks (\checked) indicate relevant results. While both RelExample and SemanticField outperform the TagPosition baseline, the results of SemanticField show more diversity for this ambiguous tag. 
The difference between (c) and (d) suggests that the results of RelExample can be diversified by varying the visual feature in use.
} \label{fig:jaguar}
\end{figure}

Concerning the influence of the media dimension, the tag + image based methods (KNN, TagVote, TagProp, TagCooccur+, TagFeature, RobustPCA, RelExample) are in general better than the tag based method (SemanticField and TagCooccur).
Fig. \ref{fig:flickr55ap} shows the per-tag retrieval performance on Flickr51. 
For 33 out of the 51 test tags, RelExample exhibits average precision higher than 0.9.
By examining the top retrieved images, we observe that the results produced by tag + image based methods and tag based methods are complementary to some extent.
For example, consider `military', one of the test tags of NUS-WIDE. RelExample retrieves images with strong visual patterns such as military vehicles, while SemanticField returns images of military personnel. Since the visual content is ignored, the results of SemanticField tend to be visually different, so making it possible to handle tags with visual ambiguity. 
This fact can be observed in Fig. \ref{fig:jaguar},  which shows the top 10 ranked images of `jaguar' by TagPosition, SemanticField, BovW + RelExample, and CNN + RelExample. 
Although their results are all correct, RelExample finds jaguar-brand cars only, while SemanticField covers both cars and animals.
However, for a complete evaluation of the capability of managing ambiguous tags, fine-grained ground truth beyond what we currently have is required.

Concerning the learning methods, TagVote consistently performs well as in the tag assignment experiment.
KNN is comparable to TagVote, due to the reason we have discussed in Section \ref{ssec:exp-tagassign}.
Given the CNN feature, the two methods even outperform their model-based variant TagProp.
Similar to the tag refinement experiment, the effectiveness of RobustPCA for tag retrieval is sensitive to the choice of visual features.
While BovW + RobustPCA is worse than the majority on Flickrt51,
the performance of CNN + RobustPCA is more stable, and performs well.
For TagFeature, its gain from using larger training data is relatively limited due to the absence of denoising.
In contrast, RelExample, by jointly using SemanticField and TagVote in its denoising component, is consistently better than TagFeature.

The performance of individual methods consistently improves as more training data is used. 
As the size of the training set increases, the performance gap between the best model-based method (RelExample) and the best instance-based method (TagVote) reduces. 
This suggests that large-scale training data diminishes the advantage of model-based methods against the relatively simple instance-based methods.

In summary, even though the performance of the methods evaluated varies over datasets, 
common patterns have been observed. 
First, the more social data for training are used the better performance is obtained.
Since the tag relevance functions are learned purely from social data without any extra manual labeling, and social data are increasingly growing, this result promises that better tag relevance functions can be learned.
Second, 
given small-scale training data,
tag + image based methods that conducts model-based learning with denoised training examples turn out to be the most effective solution,
This however comes with a price of reducing the visual diversity in the retrieval results.
Moreover, the advantage of model-based learning vanishes as more training data and the CNN feature are used,
and TagVote performs the best.

\subsection{Flickr versus ImageNet} \label{ssec:imagenet}


To address the question of whether one shall resort to an existing resource such as ImageNet for tag relevance learning,
this section presents an empirical comparison between our Flickr based training data and ImageNet.
A number of methods do not work with ImageNet or require modifications.
For instance, tag + image + user information based methods must be able to remove their dependency on user information, 
as such information is unavailable in ImageNet.
Tag co-occurrence statistics is also strongly limited, because an ImageNet example is annotated with a single label. 
Because of these limitations, we evaluate only the two best performing methods, TagVote and TagProp.
TagProp can be directly used since it comes from classic image annotation, while TagVote is slightly modified by removing the unique user constraint. 
The CNN feature is used for its superior performance against the BovW feature.

To construct a customized subset of ImageNet that fits the three test sets, 
we take ImageNet examples whose labels precisely match with the test tags. 
Notice that some test tags, e.g., `portrait' and `night', have no match, while some other tags, e.g, `car' and `dog', have more than one matches.
In particular, MIRFlickr has 2 missing tags, while the number of missing tags on Flickr51 and NUS-WIDE is 9 and 15.
For a fair comparison these missing tags are excluded from the evaluation.
Putting the remaining test tags together, 
we obtain a subset of ImageNet, containing 166 labels and over 200k images, termed ImageNet200k.

\begin{figure}[!tb]
\centering
                \includegraphics[width=0.9\textwidth]{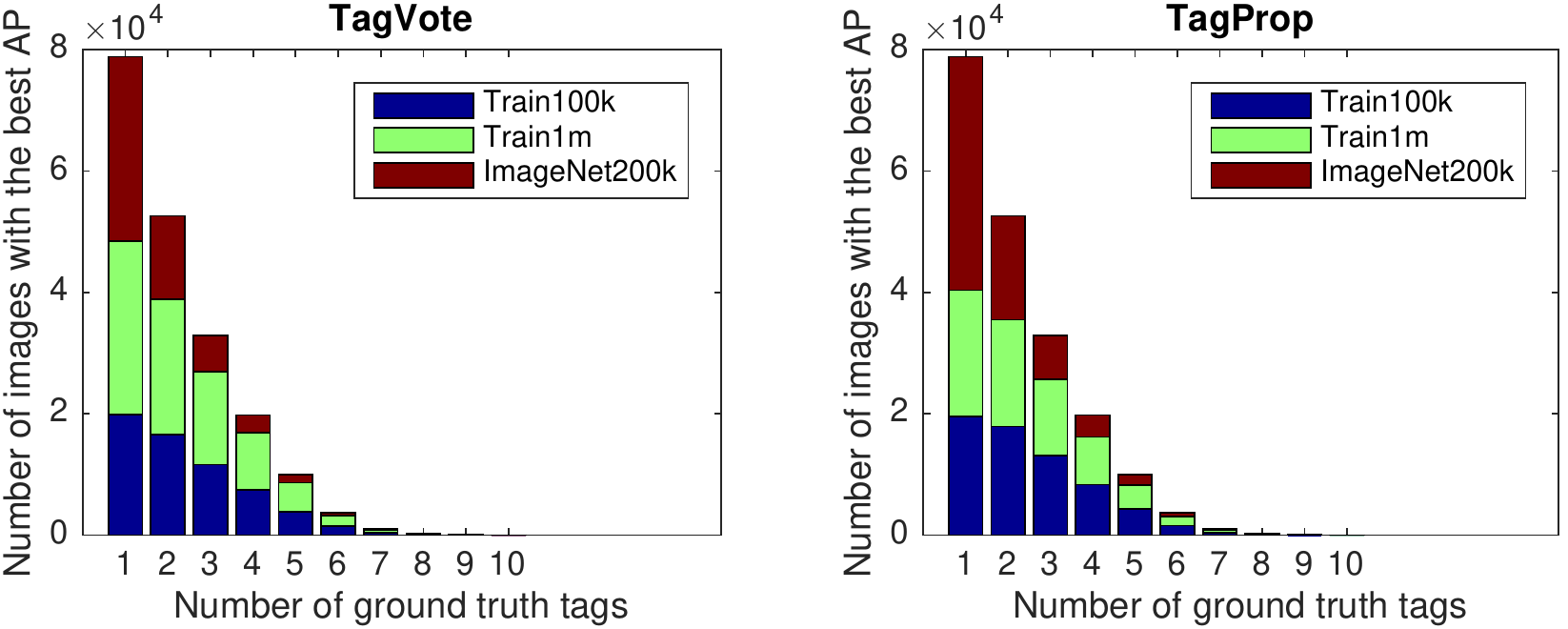}
\caption{
\textbf{Per-image comparison of TagVote/TagProp learned from different training datasets}, tested on NUS-WIDE.
Test images are grouped in terms of the number of ground truth tags. 
Within each group, the area of a colored bar is proportional to the number of images that (the method derived from) the corresponding training dataset scores the best. ImageNet200k is less effective for assigning multiple labels to an image.
}
\label{fig:miap_comparison_imagenet}
\end{figure}

The left half of Table \ref{tab:exp-imagenet} shows the performance of tag assignment.
TagVote/TagProp trained on the ImageNet data are less effective than their counterparts trained on the Flickr data. 
For a better understanding of the result, we employ the same visualization technique as used in Section \ref{ssec:exp-tagassign},
i.e., grouping the test images in terms of the number of their ground truth tags, 
and subsequently checking the performance per group.
As shown in Fig. \ref{fig:miap_comparison_imagenet},
while ImageNet200k performs better on the first group, 
i.e., images with a single relevant tag,
it is outperformed by Train100k and Train1M on the other groups.
For its single-label nature, ImageNet is less effective for assigning multiple labels to an image.

\begin{table}[tb!]
\renewcommand{\arraystretch}{1.2}
\tbl{Flickr versus ImageNet. 
Notice that the numbers on Train100k and Train1M are different from Tables \ref{tab:exp-autotag} and \ref{tab:exp-retrieval} 
due to the use of a reduced set of test tags. 
Bold values indicate top performers on a specific test set per performance metric. \label{tab:exp-imagenet}}
{\centering
\scalebox{0.9}{
\begin{tabular}{@{}l r r r r@{}}
\multicolumn{5}{c}{\textbf{Tag Assignment}}\\
\toprule
                & \multicolumn{2}{c}{\textbf{MIRFlickr}} & \multicolumn{2}{c}{\textbf{NUS-WIDE}}  \\
                \cmidrule(lr){2-3} \cmidrule(l){4-5} 
\textbf{Training Set} & TagVote & TagProp & TagVote & TagProp \\
\cmidrule{1-5} 
\textit{\textbf{MiAP scores:}} \\
\rr Train100k    & 0.377     & 0.383  & 0.392  & 0.389 \\
\rr Train1M      & 0.389     & \textbf{0.392}  & \textbf{0.414}  & 0.393 \\ [3pt]
\rr ImageNet200k  & 0.345     & 0.304  & 0.325  & 0.368 \\ [3pt]
\textit{\textbf{MAP scores:}} \\
\rr Train100k    & 0.641     & 0.647  & 0.386  & 0.405 \\
\rr Train1M      & 0.664     & \textbf{0.668}  & \textbf{0.429}  & 0.420 \\ [3pt]
\rr ImageNet200k  & 0.532     & 0.532  & 0.363  & 0.362 \\ 
\bottomrule
\end{tabular}
}
\scalebox{0.9}{
\begin{tabular}{@{}l r r r r@{}}
\multicolumn{5}{c}{\textbf{Tag Retrieval}}\\
\toprule
                & \multicolumn{2}{c}{\textbf{Flickr51}} & \multicolumn{2}{c}{\textbf{NUS-WIDE}}  \\
                \cmidrule(lr){2-3} \cmidrule(l){4-5} 
\textbf{Training Set} & TagVote & TagProp & TagVote & TagProp \\
\cmidrule{1-5} 
\textit{\textbf{MAP scores:}} \\
\rr Train100k    & 0.854     & 0.860  & 0.742  & 0.745 \\
\rr Train1M      & \textbf{0.874}     & 0.871  & 0.753  & 0.745 \\ [3pt]
\rr ImageNet200k  & 0.873     & 0.873  & \textbf{0.762}  & \textbf{0.762} \\ [3pt]
\textit{\textbf{NDCG$_{20}$ scores:}} \\
\rr Train100k    & 0.838     & 0.863  & 0.849  & 0.856 \\
\rr Train1M      & 0.894     & 0.851  & \textbf{0.891}  & 0.853 \\ [3pt]
\rr ImageNet200k  & \textbf{0.920}     & 0.898 & 0.843  & 0.847 \\ 
\bottomrule
\end{tabular}
}
}
\end{table}

For tag retrieval, as shown in the right half of Table \ref{tab:exp-imagenet}, TagVote/TagProp learned from ImageNet200k in general have higher MAP and NDCG scores than their counterparts learned from the Flickr data. 
By comparing the performance difference per concept, we find that the gain is largely contributed by a relatively small amount of concepts. 
Consider for instance TagVote + ImageNet200k and TagVote + Train1M on NUS-WIDE. 
The former outperforms the latter for 25 out of the 66 tested concepts. 
By sorting the concepts according to their absolute performance gain, the top three winning concepts of TagVote + ImageNet200k are `sand', `garden', and `rainbow', with AP gain of 0.391, 0.284, and 0.176, respectively. 
Here, the lower performance of TagVote + Train1M is largely due to the subjectiveness of social tagging. 
For instance, Flickr images labeled with `sand' tend be much more diverse, showing a wide range of things visually irrelevant to sand.
Interestingly, the top three losing concepts of TagVote + ImageNet200k are `running', `valley', and `building', with AP loss of 0.150, 0.107, and 0.090, respectively.  For these concepts, we observe that their ImageNet examples lack diversity. E.g., `running' in ImageNet200k mostly shows a person running on a track. In contrast, the subjectiveness of social tagging now has a positive effect on generating diverse training examples.

In summary, for tag assignment social media examples are a preferred resource of training data. For tag retrieval ImageNet yields better performance, yet the performance gain is largely due to a few tags where social tagging is very noisy. In such a case, controlled manual labeling seems indispensable. In contrast, with clever tag relevance learning algorithms, social training data demonstrate competitive or even better performance for many of the tested tags. 
Nevertheless, where the boundary between the two cases is precisely located remains unexplored.


\section{Conclusions and Perspectives} \label{sec:conclusions}
\subsection{Concluding remarks} \label{ssec:conclusions}

This paper presents a survey on image tag assignment, refinement and retrieval, with the hope of illustrating connections and difference between the many methods and their applicabilities, and consequently helping the interested audience to either pick up an existing method or devise a method of their own given the data at hand.
As the topics are being actively studied, inevitably this survey will miss some papers. Nevertheless, it provides a unified view of many existing works, and consequently eases the effort of placing future works in a proper context, both theoretically and experimentally. 

Based on the key observation that all works rely on tag relevance learning as the common ingredient, existing works, which vary in terms of their methodologies and target tasks, have been interpreted in a unified framework. Consequently, a two-dimensional taxonomy has been developed, allowing us to structure the growing literature in light of what information a specific method exploits and how the information is leveraged in order to produce their tag relevance scores. Having established the common ground between methods, a new experimental protocol has been introduced for a head-to-head comparison between the state-of-the-art. A selected set of eleven representative works were implemented and evaluated for tag assignment, refinement, and/or retrieval. 
The evaluation justifies the state-of-the-art on the three tasks.

Concerning what media is essential for tag relevance learning, tag + image is consistently found to be better than tag alone. While the joint use of tag, image, and user information (via TensorAnalysis) demonstrates its potential on small-scale datasets, it becomes computationally prohibitive as the dataset size increases to 100k and beyond. 
Comparing the three learning strategies, instance-based and model-based methods are found to be more reliable and scalable than their transduction-based counterparts.  As model-based methods are more sensitive to the quality of social image tagging, a proper filtering strategy for refining the training media is crucial for their success. Despite their leading performance on the small training dataset, we find that the performance gain over the instance-based alternatives diminishes as more training data is used. Finally, the CNN feature used as a substitute for the BovW feature brings considerable improvements for all the tasks.

Much progress has been made. Given the current test tag set, the best-performing methods already outperform user-provided tags for tag assignment (MiAP of 0.392 versus 0.204 on MIRFlickr and 0.396 versus 0.255 on NUS-WIDE). Image retrieval using learned tag relevance also yields more accurate results compared to image retrieval using original tags (MAP of 0.881 versus 0.595 on Flickr55 and 0.738 versus 0.489 on NUS-WIDE). 
For tag assignment and tag retrieval, methods that exploit tag + image media by instance-based learning take the leading position. 
In particular, for tag assignment, TagProp and TagVote perform best. 
For tag retrieval, TagVote achieves the best overall performance. 
Methods that exploit tag + image by transduction-based learning are more suited for tag refinement.
RobustPCA is the choice for this task.
These baselines need to be compared against when one advocates a new method.

\subsection{Reflections on future work} \label{ssec:future}
Much remains to be done. Several exciting recent developments open up new opportunities for the future.  
First, employing novel deep learning based visual features is likely to boost the performance of the tag + image based methods. What is scientifically more interesting is to devise a learning strategy that is capable of jointly exploiting tag, image, and user information in a much more scalable manner than currently feasible. 
The importance of the filter component, which refines socially tagged training examples in advance to learning, is underestimated. As denoising often comes with the price of reducing visual diversity, more research attention is required to understand what an acceptable level of noise shall be for learning tag relevance.
Having a number of collaboratively labeled resources publicly available, research on joint exploration of social data and these resources is important. 
This connects to the most fundamental aspect of content-based image retrieval in the context of sharing and tagging within social media platforms: to what extent a social tag can be trusted remains open. 
Image retrieval by multi-tag query is another important yet largely unexplored problem. For a query of two tags, it is suggested to view the two tags as a single bi-gram tag \cite{tmm12-li,mm2012-nie,mm2013-borth}, which is found to be superior to late fusion of individual tag scores. Nonetheless, due to the increasing sparseness of n-grams, how to effectively answer generic queries of more than two tag is challenging.
Test tags in the current benchmark sets were picked based on availability. 
It would be relevant to study what motivates people to search images on social media platforms and how the search is conducted.
We have not seen any quantitative study in this direction. 
Last but not least, fine-grained ground truth that enables us to evaluate various tag relevance learning methods for answering ambiguous tags is currently missing.

\medskip

``One way to resolve the semantic gap comes from sources outside the image ...'', Smeulders \etal wrote at the end of their seminal paper  
\cite{cbir-tpami00}. 
While what such sources would be was mostly unknown by that time, 
it is now becoming evident that the many images shared and tagged in social media platforms are promising to resolve the semantic gap.
By adding new relevant tags, refining the existing ones or directly addressing retrieval, 
the access to the semantics of the visual content has been much improved.
This is achieved only when appropriate care is taken to attack the unreliability of social tagging.

\begin{acks}
The authors thank Dr. Jitao Sang for providing the TensorAnalysis results, and Dr. Meng Wang and Dr. Yue Gao for making the Flickr51 dataset available for this survey.
\end{acks}

\bibliographystyle{ACM-Reference-Format-Journals}
\bibliography{review}


\end{document}